\documentclass[fleqn,usenatbib]{mnras}

\usepackage{newtxtext,newtxmath}

\usepackage[T1]{fontenc}

\DeclareRobustCommand{\VAN}[3]{#2}
\let\VANthebibliography\thebibliography
\def\thebibliography{\DeclareRobustCommand{\VAN}[3]{##3}\VANthebibliography}

\usepackage{graphicx}	
\usepackage{amsmath}	
\usepackage{amssymb}	

\usepackage[dvipsnames,xcdraw]{xcolor}

\newcommand{\kms}{\mbox{$\>{\rm km\, s^{-1}}$}}

\newcommand{\kpc}{\mbox{$\>{\rm kpc}$}} 
\newcommand{\pc}{\mbox{$\>{\rm pc}$}} 
\newcommand{\Gyr}{\mbox{$\>{\rm Gyr}$}}

\newcommand{\Msun}{\>{\rm M_{\odot}}}

\newcommand\degrees{^\circ}
\newcommand{\avg}[1]{\mbox{$\left<{#1}\right>$}}

\def\eg{{\it e.g.}}

\def\ie{{\it i.e.}}

\def\araa{ARAA}
\def\aj{AJ}
\def\apj{ApJ}
\def\apjl{ApJ}
\def\apjs{ApJS}
\def\aap{A\&A}
\def\mnras{MNRAS}
\def\astl{AstL} 
\def\pasp{PASP} 
\def\nat{Nature}

\def\aaps{A\&AS}

\long\def\Ignore#1{\relax}

\title[Bar growth revealed by flat density profiles]{The secular growth of bars revealed by flat (peak + shoulders) density profiles}

\author[Anderson et al.]{Stuart Robert Anderson$^1$\thanks{E-mail: SRAnderson1@uclan.ac.uk}, Victor P. Debattista$^1$, Peter Erwin$^2$, David J. Liddicott$^1$,
\newauthor {Nathan Deg$^3$ and Leandro {Beraldo e Silva}$^{1,4}$}
\\
$^{1}$ Jeremiah Horrocks Institute, University of Central
  Lancashire, Preston, PR1 2HE, UK \\
$^{2}$ Max Planck Institut f\"ur Extraterrestrische Physik, Giessenbachstrasse, D-85748 Garching, Germany \\
$^{3}$ Department of Physics, Engineering Physics, and Astronomy, Queen's University, Kingston, ON, K7L 3N6, Canada \\
$^4$ Department of Astronomy, University of Michigan, 1085 S. University Ave., Ann Arbor, MI 48109, USA \\
}

\date{Accepted XXX. Received YYY; in original form ZZZ}

\pubyear{2021}

\begin{document}
\label{firstpage}
\pagerange{\pageref{firstpage}--\pageref{lastpage}}
\maketitle

\begin{abstract}
The major-axis density profiles of bars are known to be either exponential or `flat'.  We develop an automated non-parametric algorithm to detect flat profiles and apply it to a suite of simulations (with and without gas).  We demonstrate that flat profiles are a manifestation of a bar’s secular growth, producing a ‘shoulder’ region (an overdensity above an exponential) in its outskirts. Shoulders are not present when bars form, but develop as the bar grows. If the bar does not grow, shoulders do not form. Shoulders are often accompanied by box/peanut bulges, but develop separately from them and are independent tracers of a bar's growth. They can be observed at a wide range of viewing orientations with only their slope varying significantly with inclination. We present evidence that shoulders are produced by looped x$_1$ orbits.  Since the growth rate of the bar moderately correlates with the growth rate of the shoulder strength, these orbits are probably recently trapped. Shoulders therefore are evidence of bar growth.  The properties of the shoulders do not, however, establish the age of a bar, because secondary buckling or strong spirals may destroy shoulders, and also because shoulders do not form if the bar does not grow much. In particular, our results show that an exponential profile is not necessarily an indication of a young bar.
\end{abstract}




\section{Introduction}
\label{s:introduction}

Bars are found in $\sim70\%$ of nearby disc galaxies \citep[e.g.][]{eskridge_etal_00, menendez-delmestre+07, Erwin_2018} and are major drivers of the evolution of galactic discs, redistributing energy, angular momentum and mass \citep[e.g.][]{weinberg85, sellwood_wilkinson93, debattista_sellwood00, athanassoula03, kormendy_kennicutt04, debattista+06, diaz-Garcia2016}.  Understanding their formation and structure is therefore crucial to understanding galactic evolution. \citet{elm_elm_85} studied surface photometry of 15 barred spiral galaxies and noted two types of bars -- those whose surface-brightness profiles along the bar major axis was `flatter than' the profile outside the bar radius, and those whose profile was a steep exponential, noting that flat bars tended to be longer and stronger than exponential bars. This was confirmed by \citet{elm_etal_96} in 19 barred galaxies. Instead \citet{seigar_james_98} found no such correlation in 24 `strongly barred' galaxies, although they considered a bar profile as flat only if the profile was constant with radius. The major axis profiles of bars are now generally grouped into exponential or flat with the latter having an overall shoulder-like shape.

Variously described as `the flat part of the bar', `humps', `bumps', `ledges', `plateaux' or `shoulders', this phenomenon is a common morphological feature of many barred galaxies along the major axis. It consists of an exponential inner part of the surface density profile, followed by a section with a much shallower gradient (not necessarily completely flat), then a much steeper downward bend to a steep exponential profile once more, often beyond the bar radius, further out in the disc.

\citet{elmegreen_96}, \citet{elm_elm_96}, and \citet{Regan_Elme_97} noted that early-type barred galaxies were more likely to have flat bars and isophotal twists than late-type galaxies, associating the twists with the presence of an inner Lindblad resonance (ILR). As part of their analysis of the bar fraction and characteristics in 2,106 disc galaxies, \citet{aguerri+2009} modelled three types of bars, one of which was the flat type. In their study of 46 galaxies, \citet{elm_etal_2011} also noted a tendency of early-type galaxies to have flatter bar profiles. \citet{kim+2015} studied 144 face-on ($i<60\degrees{}$) barred galaxies, and found more massive and bulge-dominated galaxies had flat bars, whereas less massive galaxies had exponential profiles. 

Evidence for flat bar profiles has also been found in edge-on galaxies. \citet{tsikoudi_80} noted a `hump' in the inner portion of the major axis $B$-band luminosity profile of NGC~4111 and plateaux in that of NGC~4762, attributing them to a lens structure. \citet{Wakamatsu_Hamabe_84} also noted a `hump' in the profile of NGC~4762 parallel to the major axis. \citet{DOnofrio_etal_99} examined the radial density profile of NGC~128 and noted `humps' which became less pronounced at higher height. \citet{lutticke+00} reached a similar conclusion for a sample of 60 edge-on galaxies using NIR observations, and quantified the humps using profile gradient measurements. In their study of 30 edge-on barred galaxies, \citet{bureau+06} found that 78\% of those with a BP bulge had a flat intermediate region in the major-axis brightness profile.

Shoulder-like profiles have been seen in simulations \citep[e.g.][]{schwarz_84, sparke_sellwood_87, combes+90, athanassoula_beaton06}. \citet{combes_elm_93} used \textit{N}-body simulations to study bar formation and pattern speeds, attributing flat bars to the presence of an ILR. \citet{Noguchi_96} pointed out `shoulders' appearing in the surface density profile along the bar major axis, at the ends of the bar, in his simulations of bars in tidally interacting galaxies. In their study of \textit{N}-body simulations viewed edge-on, \citet{bureau_athanassoula05} noted plateaux in the major axis surface brightness profiles, and that they grow in time as the bar lengthens; they considered that the plateaux trace or are signatures of the bar (but see \citet{val_kly_03} where flat profiles were only seen in the late stages of evolution and only for strong bars).

Clearly then, many past studies have referred to shoulders within bars but, to our knowledge, no study to date has focused on this feature in its own right in simulations. In this study, we use simulations to examine the phenomenon, and to gain insight into the mechanism by which shoulders form. We develop an automated algorithm (the shoulder recognition algorithm, hereafter the \textsc{SRA}) to identify a shoulder profile in an unsupervised fashion and explore the phenomenon quantitatively. We have run it against 1,319 profiles in 16 \textit{N}-body simulation models and one pure star-forming model.

This paper is organised as follows. In Section~\ref{s:shoulder_defn_recog} we introduce our definition of the shoulder and detail the methods of shoulder identification and quantification. In Section~\ref{s:models} we describe the models and in Section~\ref{s:results} we use our methods to analyse the formation and evolution of the shoulders. In Section~\ref{s:dissolution} we examine how shoulders dissolve. In Section~\ref{s:implications_for_observations} we discuss implications for observations. In Section~\ref{s:orbits} we examine evidence for orbital support of the shoulders, and we discuss and summarise in Section~\ref{s:discussion}.

\section{Shoulder definition and recognition}
\label{s:shoulder_defn_recog}

\subsection{Shoulder definition}
\label{ss:shoulder_definition}

In many real galaxies, the bar's major-axis surface-density profile has a multi-part structure. The innermost part of the profile is steep and (often) quasi-exponential. Beyond a certain radius, it flattens to a shallower profile (an `up-bending' transition) before turning to a steeper slope (a `down-bending' transition) further out. Finally, the steep outer profile sometimes becomes shallower again at or just beyond the end of the bar as it transitions to the disc profile (another `up-bending' transition). We define the \textit{shoulder} as the combination of the middle two regions: the shallow part of the profile plus the steep falloff, which together are the outermost part of the bar. Such a profile therefore exhibits a central peak, followed by the shoulder (`peak + shoulders', see Erwin et al., \textit{in preparation}). We argue that profiles of this type are essentially the same as the `flat' profiles identified by \citet{elm_elm_85}; we remind the reader that the slope of the shallow region need not actually be close to zero.

\begin{figure}
  \includegraphics[width=\hsize]{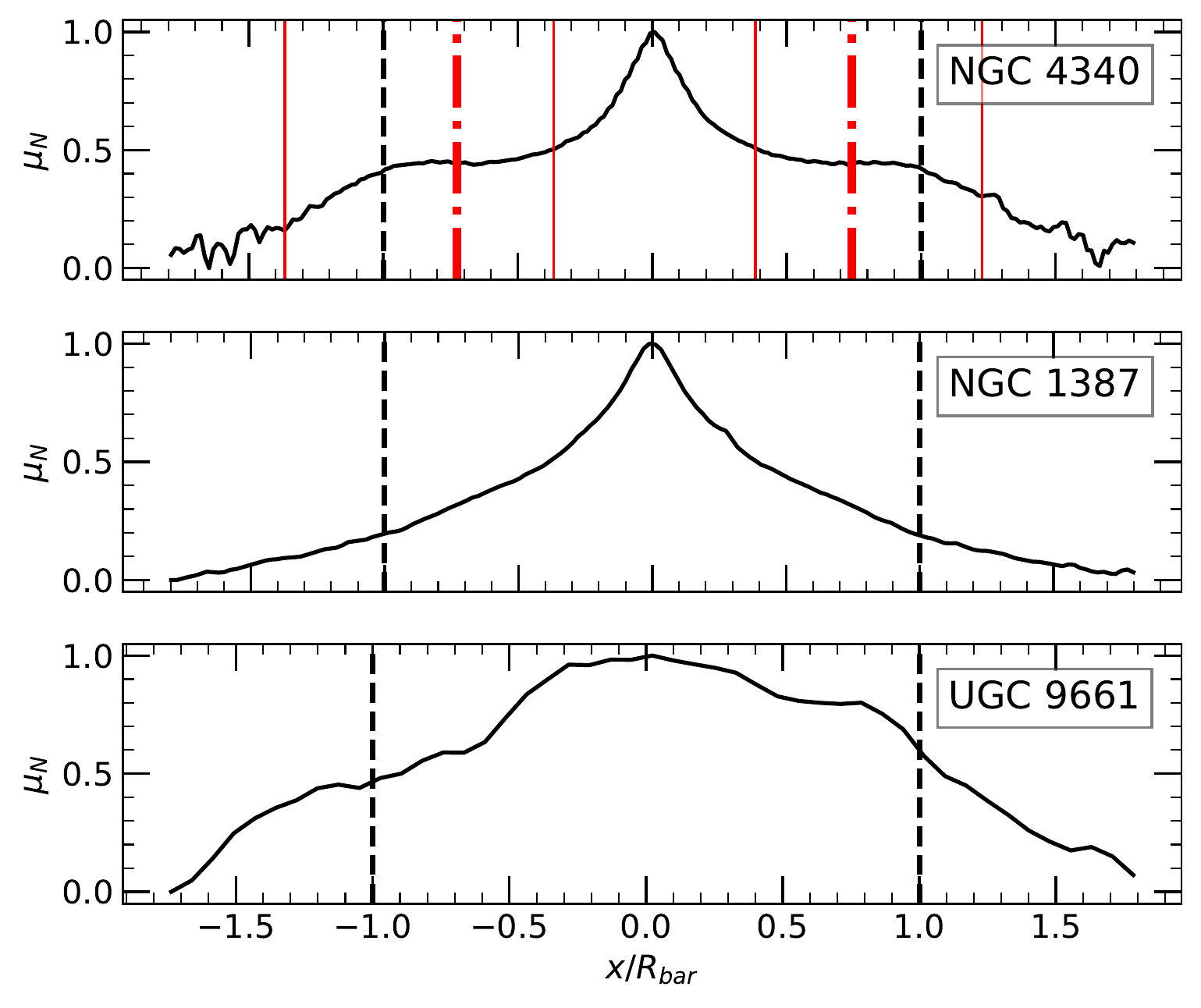}
  \caption{The normalised major axis brightness profiles (observed, not deprojected) in the \textit{Spitzer} $3.6\micron$ band for NGC~4340, NGC~1387 and UGC~9661. The black vertical dashed lines represent the bar radial extent. The shoulder recognition algorithm (see text) detects shoulders only in NGC~4340, with the thick red dot dash lines representing the centre of the clavicle and the red solid vertical lines representing the boundaries of the shoulder.}
  \label{fig:Three_real_galaxies}
\end{figure}

To illustrate this, Fig.~\ref{fig:Three_real_galaxies} shows the major axis profile of three galaxies. \textit{Spitzer} IRAC1 (3.6\micron) profiles of NGC~1387 and NGC~4340 were retrieved from the \textit{Spitzer} archive (PI K. Sheth, Program ID 10043); the image of UGC~9661 came from the \textit{Spitzer} Survey of Stellar Structure in Galaxies \citep[S$^4$G;][]{sheth+2010}. Only NGC~4340 has the shoulder profile described above, NGC~1387 has a convex profile, and UGC~9661's profile is rather noisy. The red vertical solid lines represent the inner and outer boundaries of the shoulder as found by \textsc{SRA} we have developed, which is described below. The red vertical dot dashed lines represent the centre of what we term the \textit{clavicle}\footnote{In human anatomy, the \emph{clavicle} is the collarbone, connecting the shoulder blade and the breastbone. We use it here to denote the flattest part of the shoulder structure.}, the centre of the flat portion of the shoulder, again as found by the \textsc{SRA}.

In the next subsection we describe the SRA in detail. Readers not interested in these details can skip to Section~\ref{s:models} which describes the models.

\subsection{Shoulder recognition algorithm}
\label{ss:shoulder_recog_algo}

We develop a shoulder recognition algorithm that is automatic, responsive to the signal-to-noise ratio of the underlying image (or possibly variations caused by dust and/or star formation), has tunable parameters that enable it to match by-eye detections, and provides a natural quantification of shoulder parameters. Our method is equally suited to simulations and observations but in the description which follows, we focus on its application to simulations. A list of symbols we use in the quantitative portions of this work and their meanings are given in Table~\ref{tab:symbols}.

Rather than make any \textit{a priori} assumptions about parametric profile components (for example by fitting multi-parameter exponentials or S\'ersic profiles), we use a non-parametric approach. This has the advantages of being less subjective, and requiring no visual inspection to determine fitting ranges. It can also be relatively easily automated.

\begin{table}
\centering
\caption{\label{tab:symbols}Key to symbols used in the paper.}

\begin{tabular}{ll}

\hline
Symbol & Meaning\\
\hline                                                                                           
$a$ & Slope at the clavicle centre\\
& (dimensionless, point [1] in Fig.~\ref{fig:quantification})\\

$A_{\mathrm{bar}}$ & Bar strength, calculated via the $m=2$\\
& Fourier amplitude\\

$A_{\mathrm{buck}}$ & Bar buckling amplitude\\

$\mathcal{B}$ & Strength of the BP bulge\\

$R_{\mathrm{bar}}$ & Bar radial extent\\

$R_{\mathrm{sh}}$ & Outer edge of the shoulder (point [4] in Fig.~\ref{fig:quantification})\\

$R_{\mathrm{clav,in}}$ & Inner edge of the clavicle and hence the shoulder\\ & (point [2] in Fig.~\ref{fig:quantification})\\

$\mathcal{R}_c$ & Radius of curvature of the smoothed,\\
& normalised logarithmic major axis surface\\
& density profile\\

$R_{\mathrm{BP}}$ & Radial extent of the BP bulge\\

$\Sigma_{\bigstar}$ & Stellar surface density\\

$\mathcal{S}$ & Strength of the shoulder, the fractional excess\\
& mass it contains\\

$\mathcal{Z}_{\mathrm{global}}$ & Global median height, normalised to its\\
& value at $t=0\Gyr$\\

$\rho$  & Ratio of normalised surface density at the clavicle\\ 
& to the peak density at the centre \\
& [$\log(\Sigma_\mathrm{\bigstar N, clav})/\log(\Sigma_\mathrm{\bigstar N, peak})$]\\

\hline
\end{tabular}\\
\end{table}

Since we are investigating shoulders within the bar, we measure the bar radius beforehand (see Section~\ref{ss:bar_length} for details). We then compute the logarithmic stellar surface mass density profile, $\log\Sigma_{\bigstar}(x)$, along the bar's major axis, which we take as $|y|\leq 1$ kpc (see Section~\ref{ss:bar_major_axis}). In order to compare different profiles in a uniform manner, we normalise each profile to have values between 0 and 1 and denote the result as $\log\Sigma_{\bigstar N}$. We also normalise the $x$-axis to the bar radius $R_{\mathrm{bar}}$, so both axes are dimensionless.

Our method depends on the derivatives of this profile, so smoothing is essential. Having investigated a number of smoothing techniques, we settled on using a Butterworth lowpass filter of order 2 \citep{butterworth_1930}. The algorithm then obtains derivatives of the smoothed profile by the method of central differences; a successful implementation depends on sufficient noise reduction to ensure smoothly varying derivatives, but not so much that the smoothed profile fails to faithfully follow the major profile features.

We achieved this balance for the simulations by examining several profiles, with a particular focus on model 2 of \citet{debattista+20} (see Section~\ref{s:N_body_models}) at $t=5$ Gyr. We calculated the smoothed profiles with many combinations of smoothing extent and filter order. For each combination, we calculated the root-mean-square residuals between $\log\Sigma_{\bigstar N}$ and its smoothed version, as well as the dispersion in the differences between adjacent values of the derivative $\mathrm{d}\log\Sigma_{\bigstar N}/\mathrm{d}x$, and the number of extrema in that derivative (too many extrema implying insufficient smoothing). We selected the combination which gave a reasonable balance of noise reduction (with particular attention to the number of extrema in the derivative), and a faithful representation of the overall profile shape, and validated the results by visual inspection of the smoothed versus original profiles. While this part of the analysis is subjective, it is set once and held constant throughout.

\begin{figure}
  \includegraphics[width=\hsize]{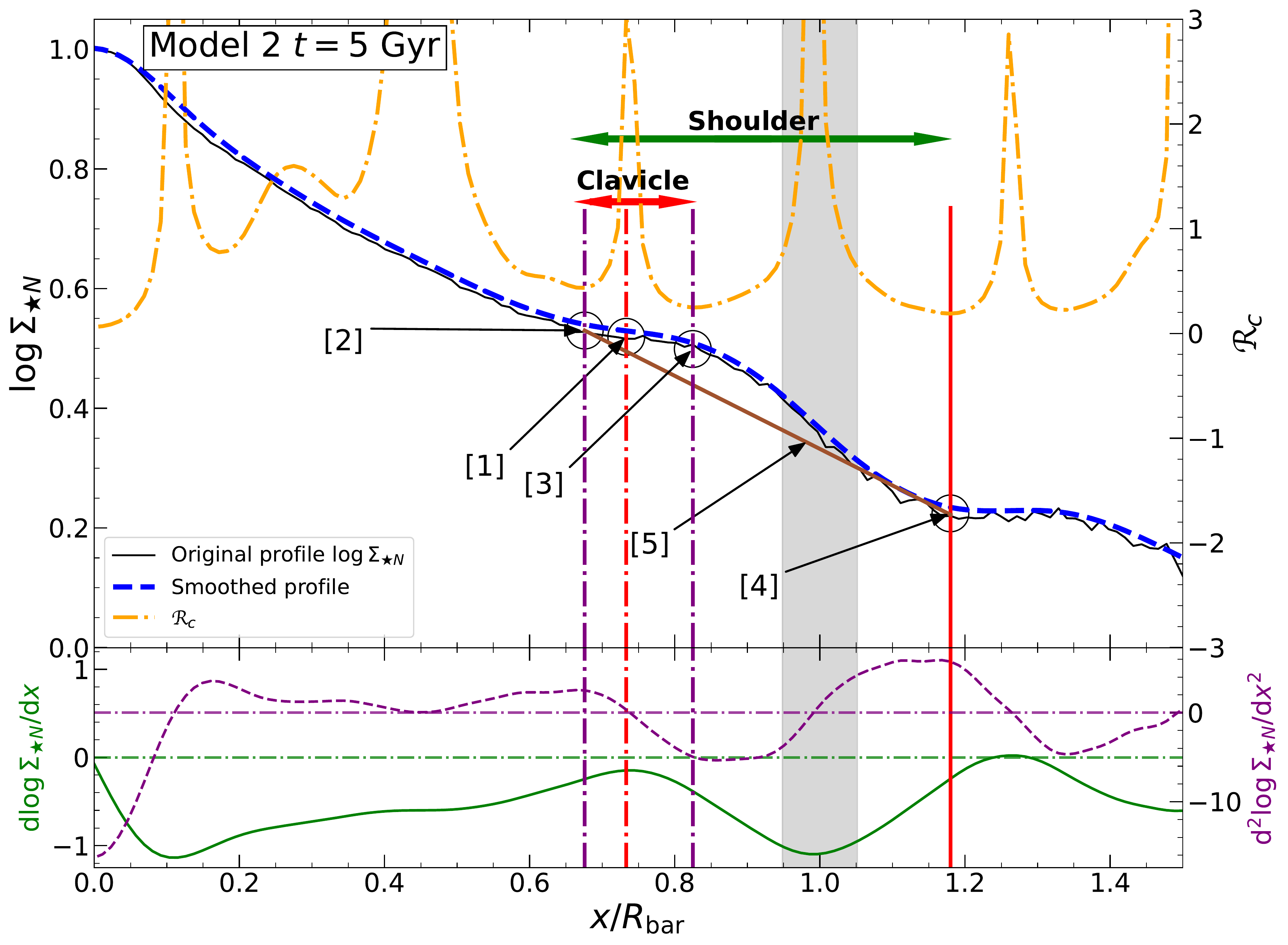}
  \caption{Normalised logarithmic density profile for model 2, at $t=5$ Gyr to illustrate the method of shoulder quantification. The $x$ axis is normalised to the bar radius, and we show only $x>0$. Upper panel: the (black) line is the original profile and the (blue) dashed line is the smoothed profile, slightly offset vertically for clarity. The radius of curvature, $\mathcal{R}_c$, is shown by the orange dot dashed curve. For clarity, we plot the derivatives in the lower panel: the green line is the first derivative of the smoothed profile, and the dashed purple line is the second derivative. The horizontal dot dash lines marks where the derivatives are zero. The permissible range of the bar radius (based on its uncertainty, see Section~\ref{ss:bar_length}), centered around $x/R_{\mathrm{bar}}=1$, is shown by the grey area. Key locations are numbered in the natural order in which they are calculated, and are described in the text.}
  \label{fig:quantification}
\end{figure}

We use the first derivative of the smoothed profile, $\mathrm{d}\log\Sigma_{\bigstar N}/\mathrm{d}x$, to test for the presence of shoulders \citep[see also][]{lutticke+00}. The process is illustrated in Fig.~\ref{fig:quantification}. For clarity, we show the profile for $x>0$, but the procedure is similar for $x<0$. The bar radius and its error are calculated using Fourier analysis of the stellar surface density projected onto the $(x, y)$-plane, described in detail in Section~\ref{ss:bar_length}.

We find all extrema of the slope $\mathrm{d}\log\Sigma_{\bigstar N}/\mathrm{d}x$ which are within the bar (i.e., at $|x| < R_\mathrm{bar}$) and at distances from the centre $> 0.2 R_\mathrm{bar}$ \citep[we impose the latter condition as we do not wish to mix shoulders with density features associated with a BP bulge, see][]{erwin_debattista16}. We then identify the flattest such extremum -- that is, the one with slope closest to zero (this marks the flattest part of the profile). Finally, we classify the profile as having a shoulder if that extremum is sufficiently flat: we define this as having a slope $< T$ (for $x < 0$) or $>-T$ (for $x > 0$). After visual inspection of many profiles, we settled on $T = 0.4$ as a reasonable threshold. Hence we do not find shoulders if (i) there is no bar or (ii) there is no region of the profile meeting the flatness criteria with respect to $T$. Note that we consider a profile which turns beyond zero slope with an up-bend to be a shoulder profile (an example can be seen in the top left panel of Fig.~\ref{fig:smoothing}).

If we do identify a shoulder, we name this minimum point in $|\mathrm{d}\log\Sigma_{\bigstar N}/\mathrm{d}x|$ as the centre of the clavicle (point [1] in Fig.~\ref{fig:quantification}). We denote the slope at the clavicle centre by $a$ -- this represents the `flatness' of the shoulder.

This analysis is repeated for $x<0$. By definition, both sides' clavicle centres must be located at $|x|<R_{\mathrm{bar}}$. Hence, ring-like structures with overdensities outside the bar are not considered shoulders. If we do not detect a clavicle on both sides of $x=0$, the \textsc{SRA} deems the profile to have no shoulders. So we only recognize shoulder pairs. Moreover, since we recognize a clavicle as the point of the smallest absolute value of the slope, the algorithm does not detect more than one pair of shoulders.

If a profile has shoulders, we determine their \textit{extent}; for this we use the radius of curvature of the smoothed profile, which is defined as:
\begin{equation}
\mathcal{R}_c = \displaystyle \frac{\left[1 + (\frac{\mathrm{d}\log\Sigma_{\bigstar N}}{\mathrm{d}x})^2\right]^{\frac{3}{2}}}{\left|\frac{\mathrm{d}^2\log\Sigma_{\bigstar N}}{\mathrm{d}x^2}\right|},
\end{equation}
and is shown as the orange dot-dashed line in Fig.~\ref{fig:quantification}. The only instance in the literature we have found which uses this parameter in a similar context is \citet{lucatelli+19}, who use the curvature ($\mathcal{R}_c^{-1}$) to identify different components in the radial density profiles of observed galaxies. \citet{foyle+08} analysed density profiles, identifying break radii, and discussed the use of profile derivatives to determine them (their Appendix B). They concluded that numerical methods were `clearly promising', although they used visual estimates for simplicity.

We set the inner boundary of the clavicle to be the location of the minimum in $\mathcal{R}_c$ nearest to the centre of the clavicle, on the side closest to $x=0$, and its outer boundary to the corresponding minimum further out. We set the outer edge of the entire shoulder to be at the second outward minimum in $\mathcal{R}_c$.

In Fig.~\ref{fig:quantification} we annotate key points in the quantification process in the natural order in which they are calculated: [1] the clavicle centre (thick red dot-dashed vertical line), at the minimum of $\mathrm{d}\log\Sigma_{\bigstar N}/\mathrm{d}x$ which is greater than $-T$, between $x=0$ and the bar radius; the slope at this point ($a$) is greater than $-T$ and so the \textsc{SRA} recognizes a shoulder; the closer $a$ is to zero, the flatter the shoulder; [2] the inner boundary of the clavicle, where $\mathcal{R}_c$ reaches its first inward minimum from the clavicle centre (thick purple dot-dashed vertical line); we define the inner boundary of the entire shoulder to be located here also; [3] the outer boundary of the clavicle, where $\mathcal{R}_c$ reaches its first outwards minimum from the clavicle centre (thick purple dot-dashed vertical line); [4] the outer edge of the entire shoulder, where $\mathcal{R}_c$ reaches its second outward minimum from the clavicle centre (thin red vertical line); we denote this as $R_{\mathrm{sh}}$; [5] a simple linear extension of the inner profile connecting inner and outer boundaries of the shoulder, constructed to estimate the excess mass contained within the shoulder (see below); we consider [4] to be the point at which the profile has reached the value where it would have been, had the shoulder been absent.

Based on these definitions, the extent of the clavicle is shown by the thick red double-headed horizontal arrow, and that of the shoulder by its green counterpart. Very thin shoulders (where the clavicle width $\leq 0.05L_\mathrm{{bar}}$) are rejected on the basis that these are likely to be local transient density perturbations rather than the extended morphological feature we are seeking.

We quantify shoulder `strength' via the `excess mass' in the shoulder. We define this quantity by extending a line from the shoulder’s inner to its outer boundary (line [5] in Fig. 2) and treating this as a notional `original' profile; the difference between the actual shoulder profile and this exponential is the `excess'. Denoting the mass along the original profile between these points as $m_o$, and the mass along the line as $m_l$, we calculate the excess mass as $m_e=m_o-m_l$, which we can then normalise to the original mass between these points. Hence our measure of shoulder strength is the dimensionless $\mathcal{S} = m_e/m_o$. Stronger shoulders have higher fractional excess mass.
 
We define the errors on all calculated parameters as half the difference between the values for $x<0$ and $x>0$. As a proof of concept, we have run the \textsc{SRA} against the three bar major axis profiles shown in Fig.~\ref{fig:Three_real_galaxies}. The \textsc{SRA} correctly identifies the shoulders in NGC~4340, and does not recognize shoulders in the other two (NGC~1387 does not pass the first derivative slope threshold, and the shallowest first derivative slope for $x<0$ in NGC~9661 is outside the bar).

\section{The models}
\label{s:models}
        
The majority of our models have been presented elsewhere; for the sake of concision we refer the reader to earlier papers in such cases. We use the same naming conventions for ease of reference.

\subsection{\textit{N}-body models}
\label{s:N_body_models}



Our first set of models are pure $N$-body models with no gas or star formation. Several of these have been published already.  From \citet{debattista+20}, we use models 2, 3, 4 and 5, which are baryon-dominated pure disc Milky Way-like models.  From the same paper we also use the thin$+$thick disc model T1, as well as the dark matter-dominated model HD2. From \citet{debattista+17} we use the thin$+$thick disc model T4. 

We include an unpublished thin$+$thick disc model, T6.  This is similar to T1 in that it has a thin and a thick disc of equal mass, both having scale length $R_d = 2.4\kpc$. The main differences are in the geometric and kinematic parameters of the two discs.  T6 has an initially thicker thick disc, with scale height $h_z = 900\pc$ (versus $h_z = 400\pc$ in T1), a central velocity dispersion $\sigma_0 = 140\kms$ (versus $\sigma_0 = 90 \kms$ in T1) which declines exponentially with a scale length $R_\sigma = 3.5\kpc$ (versus $R_\sigma = 2.5 \kpc$ in T1).  The thin disc differs from that in T1 by being thicker, with $h_z = 300\pc$ (versus $h_z = 100\pc$ for the thin disc in T1).

We also include two unpublished bulge$+$disc models. These have been constructed using {\sc GalactICs} \citep{kuijken_dubinski95, widrow_dubinski05}.  They are based on the Milky Way-like models in \citet{widrow+08}, which are described also in \citet{hartmann+14}. 
The profiles of the bulges are given by %
\begin{equation}
\rho = \rho_0 \left(\frac{R}{R_b}\right)^{-p} e^{-b(R/R_b)^{1/n}}
\end{equation}
where $b$ is always set such that $R_b$ is the half-mass (effective) radius. In projection, this results in a S\'ersic profile.

Model CB1 is based on the Toomre-$Q = 1.75$, $X = 3.5$ model of \citet{widrow+08}, with a more compact and more massive bulge.  We use $n=1$, $p = 0.44$, and $R_b = 300\pc$.  The density scale $\rho_0$ is not set directly but is set via the scale velocity \citep[see][]{hartmann+14}, which we set to $\sigma_b = 350\kms$. The comparable model in \citet{widrow+08} has $n = 0.85$, $p = 0.36$, and $R_b = 670\pc$, with scale velocity $\sigma_b = 215 \kms$.
Model CB1 is comprised of 2 million disc particles, 0.4 million bulge particles and 1 million dark matter particles.

Model CB2  is based on the Toomre-$Q = 1.25$, $X = 2.5$ model of \citet{widrow+08}, with another compact bulge, having $n = 1$, $p = 0.44$, $R_b = 300\pc$ and $\sigma_b = 400\kms$. In contrast the comparable model in \citet{widrow+08} has $n = 0.85$, $p = 0.44$, and $R_b = 580\pc$, with scale velocity $\sigma_b = 240 \kms$.
Model CB2 is comprised of 0.9 million disc particles, 0.2 million bulge particles and 1 million dark matter particles.

Model PB1 is also a new model, which is comprised of a pseudobulge and an exponential disc built using the version of \textsc{GalactICS} presented in \citet{deg+2019}.  The halo of model PB1 is a Hernquist model, defined using the \textsc{GalactICS} parameters $\sigma_{h}=550\kms$, $R_{h}=30\kpc$, $\alpha=1$, $\beta=4$.  The main disc has $M_{d1}=4.31\times10^{10} \Msun$, $R_{d1}=2.67\kpc$, and $z_{d1}=0.35 \kpc$ while the pseudobulge is another exponential disc with $M_{d2}=4.77\times10^{9} \Msun$, $R_{d2}=0.26 \kpc$, and $z_{d2}=0.23 \kpc$. Model PB1 is comprised of 2 million disc particles, 200,000 particles in the pseudo-bulge disc and 15 million dark matter particles.

The new model SD1 is built using a modified version of the \textsc{GalactICS} code that generates collisionless discs using a S\'ersic surface density profile:
\begin{equation}
 \Sigma_{d}=\frac{M_{d}}{2\pi n R_{d}^{2}\Gamma(2n) }e^{-(R/R_{d})^{1/n}}~,
\end{equation}
where $M_{d}$ is the disc mass, $R_{d}$ is the scale length, $n$ is the S\'ersic index, and $\Gamma$ is the gamma function.
Model SD1 has $5\times10^{6}$ halo particles and $4.4\times10^{6}$ disc particles.  The halo uses the same parameters as PB1, while the S\'{e}rsic disc has $M_{d}=5.74\times10^{10} \Msun$, $R_{d}=0.265\kpc$, $z_{d}=0.25\kpc$ and $n=2.05$.

We also include two models in which we suppress most secular bar growth by setting the halo in full prograde rotation \citep{debattista_sellwood00, long+2014, collier+2018}. This results in large spin parameters of the haloes, $\lambda$, which are rare in cosmological simulations \citep{bullock_angmom+01}. Since our goal is to contrast these models with their evolving and growing versions with the more typical, unrotating haloes, we are unconcerned by the fact that fully rotating haloes are very improbable in nature. The two models for which we have done this are model 2 (which becomes 2S, with $\lambda \simeq 0.084$) and model SD1 (which becomes SD1S, with $\lambda \simeq 0.091$).

All numerical parameters in the collisionless runs are the same as those in models 2, 3, 4 and 5 which are described in \citet{debattista+20}.

\subsection{Star-forming models}
\label{s:star_forming_models}

Models PBG1 and PBG2 are built using the version of \textsc{GalactICS} released in \citet{deg+2019}, which builds equilibrium models with initially isothermal gas discs.  These two models consist of 2 stellar discs, a gas disc, and a dark matter halo.  For both models the main stellar disc has $M_{d1}= 4.31\times 10^{10} \Msun$, $R_{d1}= 2.67\kpc$, $z_{d1}= 0.32\kpc$, and an exponential disc pseudo-bulge with $M_{d2}= 5.67\times10^{9} \Msun$, $R_{d2}=0.35\kpc$, $z_{d2}=0.3\kpc$.  They both have halos with $\sigma_{h}=550\kms$, $R_{h}=30\kpc$, $\alpha=1$, and $\beta=4$.  Both models have a gas disc of mass $M_{g}= 4.79\times10^{9} \Msun$. The difference between the two models is that model PBG1 has a gas disc with scale-length $R_{g}= 2.67 \kpc$, while model PBG2 has a more extended disc, with $R_{g}=6.7\kpc$.

We also use the star-forming simulation described in \citet{cole+14}, \citet{gardner+14} and, most extensively, in \citet{debattista+17}. Uniquely, this model starts out with a gas corona but no stars at all, so its evolution is free of any potential biases that might arise from inserting a disc of stars ab initio. We adopt \citet{gardner+14}'s designation for this model: HG1.

Models PBG1 and PBG2 were evolved with {\sc changa} \citep{jetley_etal_2008, jetley_etal_2010, menon_etal_2015}. Stars form out of gas with a $5\%$ efficiency, once the cool gas density exceeds 0.1 cm$^{-3}$ in a converging flow. Thermal and chemical turbulent diffusion uses the prescription of \citet{shen_et_al_2010}, with a mixing parameter $D = 0.03$. Gas and star particles have softening of $\epsilon = 50~\pc$, including newly formed stars, while the dark matter particles have $\epsilon = 100~\pc$. Supernova feedback couples 15\% of the $10^{51}$ erg per supernova to the interstellar medium. We use a base time step of $\Delta t = 2.5$~Myr with timesteps refined such that $\delta t = \Delta t/2^n < \eta\sqrt{\epsilon/a_g}$, where we set the refinement parameter $\eta = 0.175$ and $a_g$ is the gravitational acceleration at a particle's position. We set the opening angle of the tree-code gravity calculation to $\theta = 0.7$. Gas particle timesteps also satisfy the condition $\delta t_{gas} = \eta_{courant} h/[(1 + \alpha)c+ \beta \mu_{max}]$, where $\eta_{courant} = 0.4$, $h$ is the SPH smoothing length set over the nearest 32 particles, $\alpha$ and $\beta$ are the linear and quadratic viscosity coefficients and $\mu_{max}$ is described in \citet{gasoline}.

The simulations listed above do not include all the models we have studied. We include here models exhibiting interesting behaviours which help us understand the shoulder phenomenon. We emphasise that our simulations are isolated, so each model's evolution is entirely secular.

We found no obvious differences between the collisionless models, the two models with gas and the pure star-forming model HG1, so in this study we do not compare collisionless models and models with gas. We examine model HG1 in Section~\ref{ss:star_forming_model}.

\subsection{Bar strength, length and formation time}
\label{ss:bar_length}

As we are investigating profiles within the bar, we require measurements of the bar strength and radius. Since the bar is a bisymmetric deviation from axisymmetry, we define the bar strength, $A_{\mathrm{bar}}$, as the amplitude of the $m=2$ Fourier component of the stellar particle surface density distribution, projected onto the $(x,y)$-plane. Having rotated the models so the bar major axis is oriented along the $x$-axis, we calculate the azimuthal angle (with respect to the $x$ axis) $\phi_k$ of each stellar particle, of mass $m_k$, and then calculate:
\begin{equation}
A_{\mathrm{bar}} = \left|\frac{\Sigma_k m_k e^{2i\phi_k}}{\Sigma_k m_k}\right|.
\end{equation}
Several methods have been developed for measuring bar sizes \citep[e.g.][]{agu_etal_00, athanassoula_misiriotis02, erwin05, michel_dansac+wozniak_2006}.  We adopt the bar radius as the average of two calculations: the first is the (cylindrical) radius at which the amplitude of the $m=2$ Fourier component reaches half its maximum value after the peak. The second is the (cylindrical) radius at which the phase of the $m=2$ component deviates from a constant by more than $10\degrees$, with the uncertainty as half the difference. We refer to the resulting bar radius as $R_{\mathrm{bar}}$; the resulting average uncertainty is $10\%$. We calculate the time of bar formation, $t_{\mathrm{bar}}$, as the time when $\log(A_{\mathrm{bar}})$ changes from a positively sloped line to flat -- \textit{i.e.} when the instability has saturated and formed a steady bar.

\subsection{Bar major axis}
\label{ss:bar_major_axis}

We restrict our analysis to the bar by taking a cut along its major axis, $|y|\leq 1$ kpc. To show that our choice of cut in $|y|$ does not materially alter our results, we show in Fig.~\ref{fig:y_layers} the profile for various cuts in $|y|$, ranging from 500~pc to 2~kpc for model 2. Little additional profile information is gained by increasing the width of the cut while using too thin a cut only increases the noise. We therefore adopt $|y|\leq 1$ kpc for all models in this study. We call this the `major axis cut' for ease of reference, and it always implies $|y|\leq 1$ kpc.

\begin{figure}
  \includegraphics[width=\hsize]{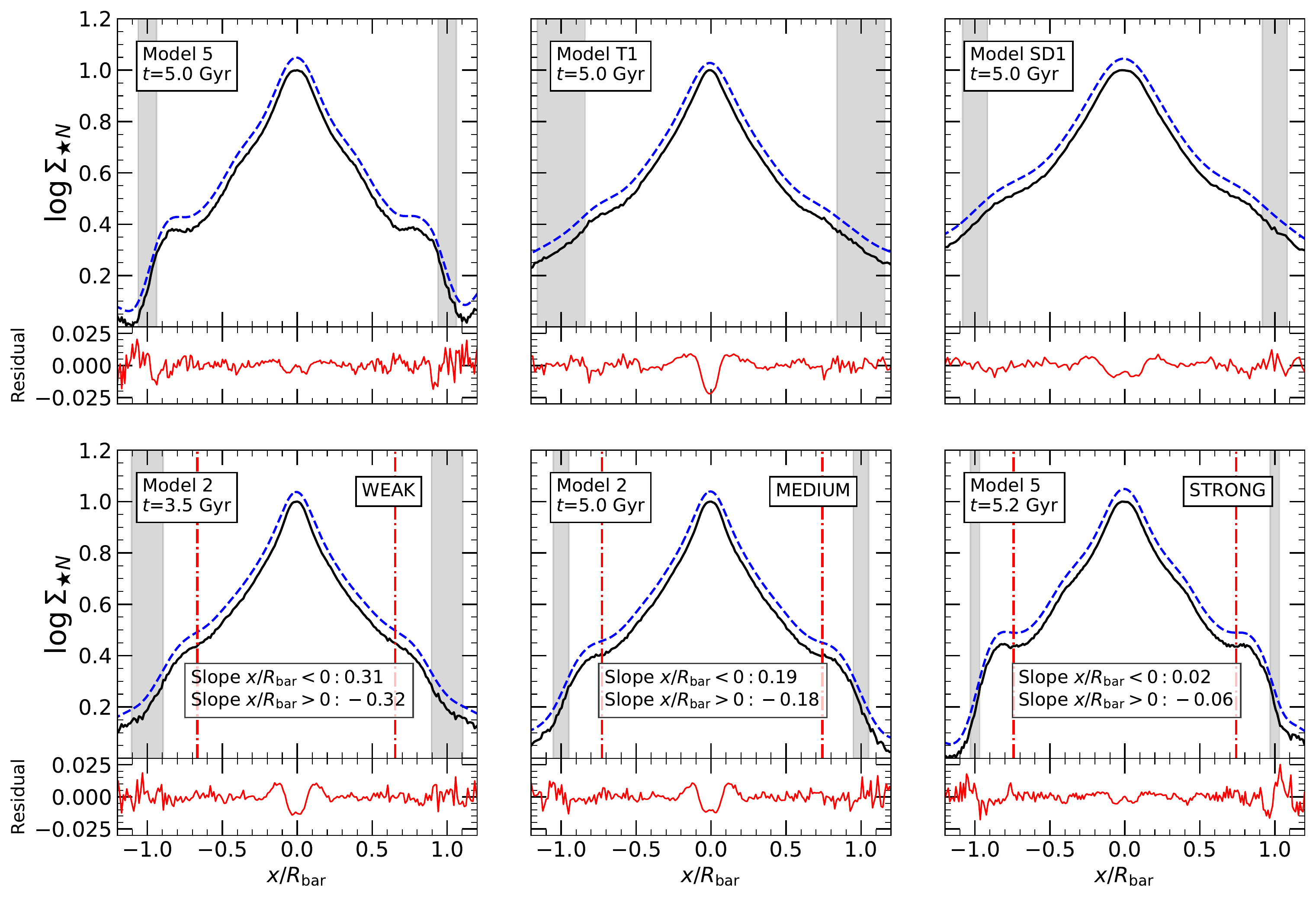}
  \caption{\textit{Upper panels}: A sample of normalised surface density profiles along the bar major axis for three models, taken with the major-axis cut ($|y|\leq 1 \kpc$). Each panel shows the original profile (black line), and the smoothed profile (dashed blue line), offset vertically for clarity. The bar radius is indicated by the grey areas. \textit{Lower panels}: Examples of weak, medium and strong shoulder profiles. The clavicle centres are indicated by the vertical red dot dashed lines, and the slopes at the clavicle are indicated in the text beneath each profile. Residuals (difference between the smoothed and original profiles) are shown beneath each panel in red.}
  \label{fig:smoothing}
\end{figure}

\begin{figure}
  \centering
  \includegraphics[width=\hsize]{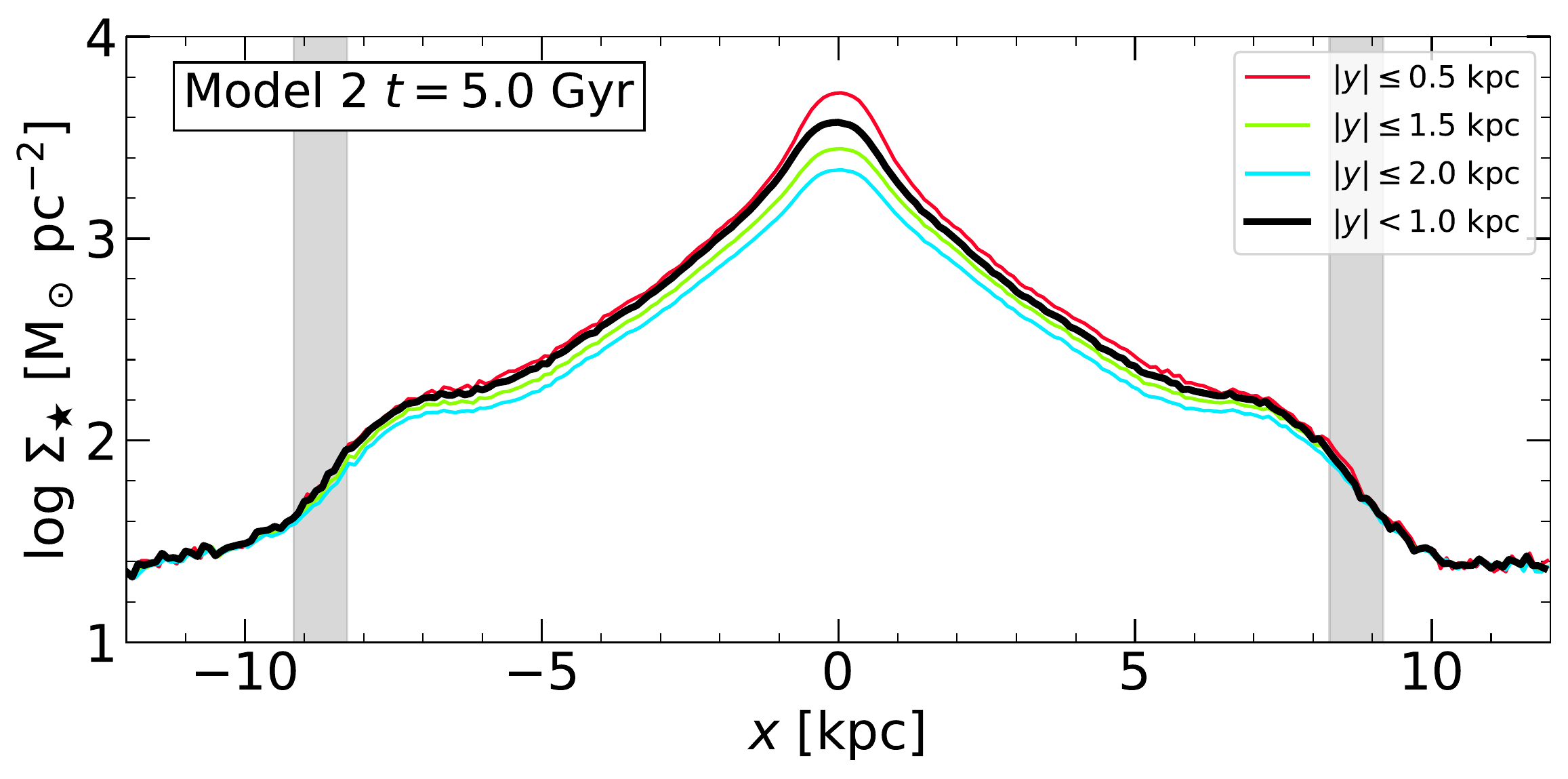}
  \caption{Logarithmic surface density profiles along the bar major axis, $x$, for various cuts in $|y|$, in model 2 at $t= 5$ Gyr. The thick black line shows the cut $|y|\leq 1$ kpc (the `major axis cut'), and is the one we use for all analyses in this study. The grey areas represent the permissible range of bar radius (see text for details).}
  \label{fig:y_layers}
\end{figure}

\subsection{Bar buckling and the box/peanut bulge}
\label{ss:buckling_BP}

The bars in some of the models suffer the buckling instability \citep{raha+91,martinez-valpuesta_shlosman04,smirnov_sotnikova19}. Buckling manifests as a deviation from vertical symmetry, followed by a rapid increase in thickness, and generally the development of a box/peanut (BP) bulge \citep{bureau_freeman99, debattista+04, athanassoula05, bureau+06} in the inner part of the bar. We define the buckling amplitude as:
\begin{equation}
A_{\mathrm{buck}} = \left|\frac{\Sigma_k z_k m_k e^{2i\phi_k}}{\Sigma_k m_k}\right|,
\end{equation}
where $z_k$ is the $z$ coordinate of the $k^{th}$ particle. $A_{\mathrm{buck}}$ has dimensions of length.

We quantify the strength of a BP bulge following the method of \citet{fragkoudi+17}: we measure the maximum of the median of absolute heights ($\lvert{z}\rvert$) for the particles in radial bins of $0.1\kpc$. Normalising by the global median of absolute heights at $t=0$ ($|z|_0$) for each model makes comparison between models possible. So we define the BP strength, with tilde representing the median, as:
\begin{equation}
\mathcal{B} = \frac{\widetilde{|z(R)|}_\mathrm{max}}{\widetilde{|z|}_{0}}.
\label{eq:B}
\end{equation}
We measure the radial extent of the BP bulge, $R_{\mathrm{BP}}$, as the radius at which $|z|$ (and therefore $\mathcal{B}$) is a maximum. As a measure of vertical heating, we denote the global median $|z|$ for a model scaled to its value at $t=0$ as $\mathcal{Z}_{\mathrm{global}}$.

\section{Results}
\label{s:results}

\subsection{Sample model profiles}
\label{s:sample_model_profiles}

Fig.~\ref{fig:smoothing} (upper panels) shows examples of the original and smoothed profiles and their residuals for a selection of models. The residuals outside the bar are generally larger than those within, but this is not a concern as we are interested in the profiles within the bar. We also find that the central regions of highly concentrated models can have larger residuals; however these inner regions are not of concern either because the shoulders do not reside there.

Judging whether a profile has shoulders is somewhat subjective; as discussed earlier, the \textsc{SRA} uses the clavicle slope as a threshold. Consider the three profiles shown in the lower panels of Fig.~\ref{fig:smoothing}. In all cases the profile slope changes at $|x/R_\mathrm{bar}| \sim 0.75$. At this point, the profile on the left has barely perceptible shoulders, the middle panel exhibits clear shoulders, while the right panel has strong shoulders. The slopes' absolute values decrease as the shoulders strengthen. Thus the slope lends itself to shoulder identification and quantification -- there is an absolute slope value above which we consider shoulders to be absent.

\begin{figure*}
  \centering
  \includegraphics[width=\hsize]{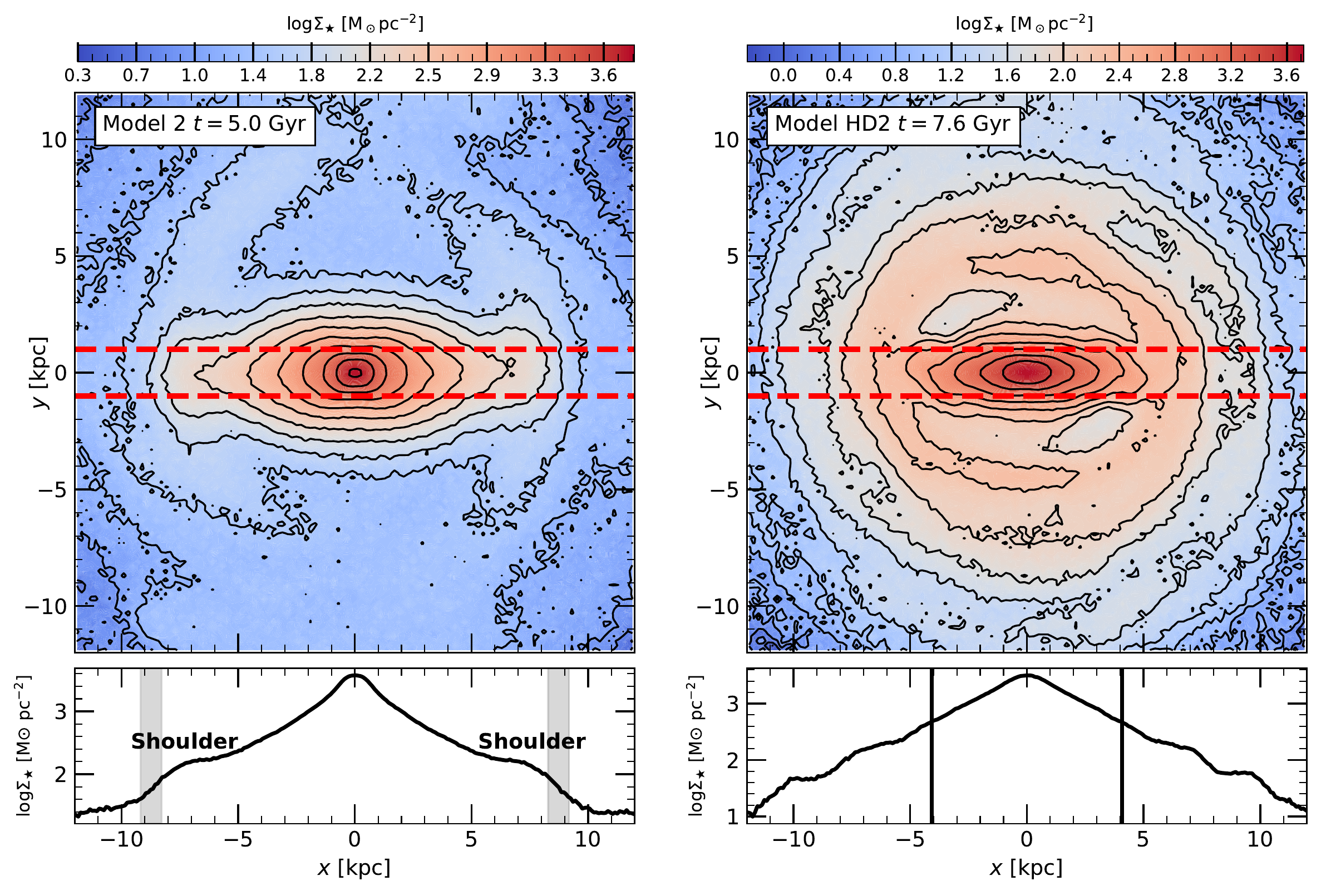}
  \caption {\textit{Upper panels}: Logarithmic stellar surface density plot in the $(x,y)$-plane for models 2 at $t = 5 \Gyr$ (left) and HD2 at $t=7.6$ Gyr (right). The red horizontal dashed lines represent $y=\pm$1 kpc.  \textit{Lower panels}: logarithmic stellar surface density profile along the bar major axis for $|y|\leq 1$ kpc, for the respective models and time steps. The grey areas represent the bar radius (vertical black lines for model HD2 as the error in bar radius is small at this time step).  In model 2, the shoulders are centred at $|x| \sim 7\kpc$. In model HD2, the ring encircling the bar at $\sim5-7$ kpc produces a flat profile similar to shoulders, but lies outside the bar and so does not qualify as a shoulder profile.}
  \label{fig:D6_HD2_overview}
\end{figure*}

As an example from our simulations, the left panels of Fig.~\ref{fig:D6_HD2_overview} show the logarithmic surface density plot in the $(x,y)$-plane for model 2 at $t = 5 \Gyr$ (the same model and time as in Fig.~\ref{fig:y_layers}). The model has a strong bar; the shoulder profile is visible at $|x|\sim 7 \kpc$ within the bar radius (indicated by the grey shaded regions), and is roughly symmetric about $x=0$. This is accompanied by an increase in the contour spacing along the major axis. The downwards bend occurs just after the end of the bar, as was found in the simulations of \cite{athanassoula_beaton06}.

Note that we consider an `up-bending' profile within the bar to be a shoulder profile. An example can be seen in the top left panel of Fig.~\ref{fig:smoothing} (model 5 at $t=5$ Gyr). The \textsc{SRA} recognizes this as a shoulder.

Outer rings are not considered shoulders. The right hand panels of Fig.~\ref{fig:D6_HD2_overview} show model HD2 at $t=7.6$ Gyr. The flat overdensities at $|x| \sim 6$ and $9$ kpc are \emph{not} shoulders, since they lie outside the bar. The surface density plot in the upper panel shows that these are \emph{rings}. Moreover, shoulders are not the same phenomena as Freeman Type II profiles \citep{freeman70, erwin+08}, down-bending disc breaks in the azimuthally averaged profiles which occur much further out in the disc than the bar.

\subsection{Bar formation and buckling}
\label{s:quantitative_results}

\begin{figure*}
  \centering
  \includegraphics[width=\hsize]{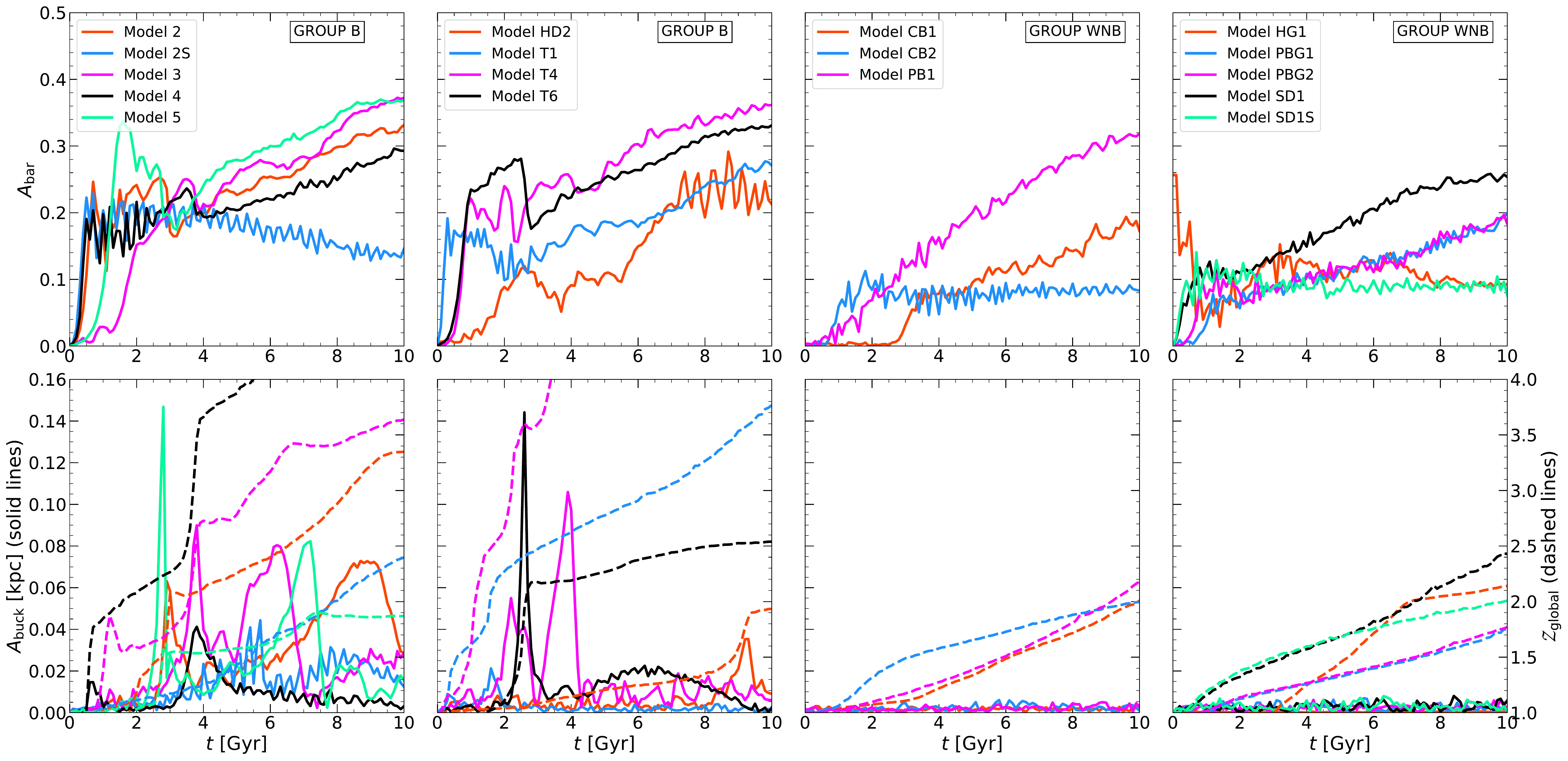}
  \caption{The global bar and buckling amplitudes $A_{\mathrm{bar}}$ and $A_{\mathrm{buck}}$ for models in groups B and WNB. Top row: $A_{\mathrm{bar}}$. Bottom row: $A_{\mathrm{buck}}$ (solid lines) and $\mathcal{Z}_{\mathrm{global}}$ (dashed lines). Note that the ordinate axis scale is the same in each panel to ease comparison. Buckling amplitudes are close to 0 in most WNB models, and $\mathcal{Z}_{\mathrm{global}}$ increases more steadily in these models than in B models.}
  \label{fig:barbuck_amp}
\end{figure*}

We start our study of the simulation suite by describing the evolution of the bars in the models.
Fig.~\ref{fig:barbuck_amp} shows the evolution of the global bar and buckling amplitudes, $A_{\mathrm{bar}}$ and $A_{\mathrm{buck}}$.
We group the simulations into those that strongly buckle (group B) and those that have weak or no buckling (group WNB). Five of the group B models (models 2, 2S, 3, 5 and T4) undergo a second major buckling. In most group WNB models, bar strength grows smoothly (top row, right two columns), but there are some with flat or declining bar strengths. We have explored other ways of grouping the models, including the bar `speed' (fast or slow as determined by $\mathcal{R}$, the ratio of corotation radius to bar radius, with a separation at $\mathcal{R} =1.4$), and heights within and outside the bar. However we were unable to find any more insightful grouping than those that buckle, and those which do not (or do so very weakly). We also checked central concentration via the $C_{28}$ parameter \citep{kent_1987_2}, defined as $5\log (R_{80}/R_{20})$, with $R_{80}$ and $R_{20}$ being the cyclindrical radii containing 80\% and 20\% of the stellar mass, respectively. A high central mass concentration suppresses buckling \citep{berentzen+07,iannuzzi_ath_2015,seo+2019}, so this is degenerate with the B/WNB grouping we adopt, with $\avg{C_{28}}=2.78 (3.78)$ for B (WNB) models at $t=0$. We therefore rely on the B and WNB groupings throughout.

Fig.~\ref{fig:barbuck_amp} also shows the evolution of $\mathcal{Z}_{\mathrm{global}}$ (bottom row, dashed line); for model HG1, we normalise this to $\widetilde{|z|}$ at $t=3\Gyr$ since at $t=0$ we have no stellar particles, and by $3\Gyr$ we have a stable bar. This ratio increases with time in both groups. Each major buckling episode in group B is accompanied by a large increase in $\mathcal{Z}_{\mathrm{global}}$, as the buckling heats the disc vertically. In contrast, in group WNB models the increase in $\mathcal{Z}_{\mathrm{global}}$ is gradual.
Table \ref{tab:models-shoulders-overview} presents an overview of some key evolutionary parameters.

\begin{table*}
\centering
\caption{\label{tab:models-shoulders-overview}Key evolutionary characteristics of the models.}

\begin{tabular}{llllllllllllllll}

\hline
Model  & Group & $t_{\mathrm{bar}}$ & $t_{\mathrm{buck}}$ & $R_{\mathrm{bar,end}}$ & $t_{\mathrm{sh}}$ & $t_{\mathrm{sh}} - t_{\mathrm{bar}}$ & $t_{\mathrm{sh}} - t_{\mathrm{buck}}$ & $\mathcal{S}_{\mathrm{max}}$ & $a_\mathrm{shallowest}$ \\
 & & [Gyr] & [Gyr] & {[}kpc{]}  & {[}Gyr{]} & {[}Gyr{]} & {[}Gyr{]} &  & \\
\hline                                                                                           
Model 2$^*$           & B  &  0.7  & 2.9 & 11.6  & 3.5   & 2.8  & 0.6  & 0.24 & -0.18 \\ 
Model 3$^*$           & B  &  2.1  & 3.8 & 11.0  & 3.1   & 1.0  & -0.7 & 0.26 & -0.12 \\ 
Model 4               & B  &  0.5  & 3.8 & 9.3   & 4.1   & 3.6  & 0.3  & 0.14 & -0.18 \\ 
Model 5$^*$           & B  &  1.6  & 2.9 & 9.7   & 3.5   & 1.9  & 0.6  & 0.35 & -0.03 \\ 
Model T6              & B  &  1.0  & 2.6 & 13.7  & -     & -    & -    & -    & - \\ 
Model HD2$\dagger$    & B  &  2.6  & 9.3 & 4.8   & 6.3   & 3.7  & -3.0 & 0.28 & 0.05 \\ 
Model T1              & B  &  0.3  & 1.6 & 10.1  & 3.8   & 3.5  & 2.2  & 0.13 & -0.2 \\ 
Model T4$^*$          & B  &  1.0  & 2.2 & 12.6  & 1.9   & 0.9  & -0.3 & 0.25 & -0.26 \\ 
Model 2S$^*$          & B &  0.5  & 5.5   & 7.4   & -     & -    & - & -    & -     \\    
\hline                                                                                        
Model HG1$\ddagger$   & WNB &  3.5  & -   & 3.0   & 6.7   & 3.2  & - & 0.08 & -0.22 \\    
Model CB1             & WNB &  3.2  & -   & 7.2   & 3.2   & 0.0  & - & 0.16 & -0.21 \\    
Model PB1             & WNB &  1.0  & -   & 7.1   & 2.3   & 1.3  & - & 0.24 & -0.04 \\    
Model CB2             & WNB &  1.1  & -   & 5.7   & -     & -    & - & -    & - \\    
Model PBG1           & WNB &  0.6  & -   & 6.2   & 3.5   & 2.9  & - & 0.13 & -0.17 \\    
Model PBG2          & WNB &  0.3  & -   & 6.3   & 0.8   & 0.0  & - & 0.12 & -0.15 \\    
Model SD1           & WNB &  0.8  & -   & 7.8   & 4.0   & 3.2  & - & 0.21 & 0.09 \\    
Model SD1S          & WNB &  0.8  & -   & 3.6   & -     & -    & - & - & - \\    

\hline
\end{tabular}\\
\begin{flushleft}
$t_{\mathrm{bar}}$: The time of bar formation, defined in Section~\ref{ss:bar_length}.\newline
$R_{\mathrm{bar, end}}$: The length of the bar at the end of the model's run.\newline
$t_{\mathrm{buck}}$: The time of the first major peak $A_{\mathrm{buck}}$.\newline
$t_{\mathrm{sh}}$: The time when persistent shoulders are first recognized by the \textsc{SRA}, disregarding transients (see Section \ref{ss:evolution}).\newline
$\mathcal{S}_{\mathrm{max}}$: The maximum strength of the shoulders in the model's run, as defined in Section~\ref{ss:shoulder_recog_algo}.\newline
$a_\mathrm{shallowest}$: The shallowest slope (\ie{} maximum flatness) of the shoulders in the model's run. The closer this figure to 0, the flatter the shoulder at its shallowest.\newline\newline
$^*$These models undergo a second major buckling, defined by a second major peak in $A_{\mathrm{buck}}$.\newline
$\dagger$This model develops a BP bulge via resonant trapping before the bar buckles -- see text.\newline
$\ddagger$The pure star-forming model.
\end{flushleft}
\end{table*}

\subsection{Formation and evolution of shoulders}
\label{ss:evolution}

\subsubsection{Shoulder detection}

\begin{figure*}
  \centering
  \includegraphics[width=\hsize]{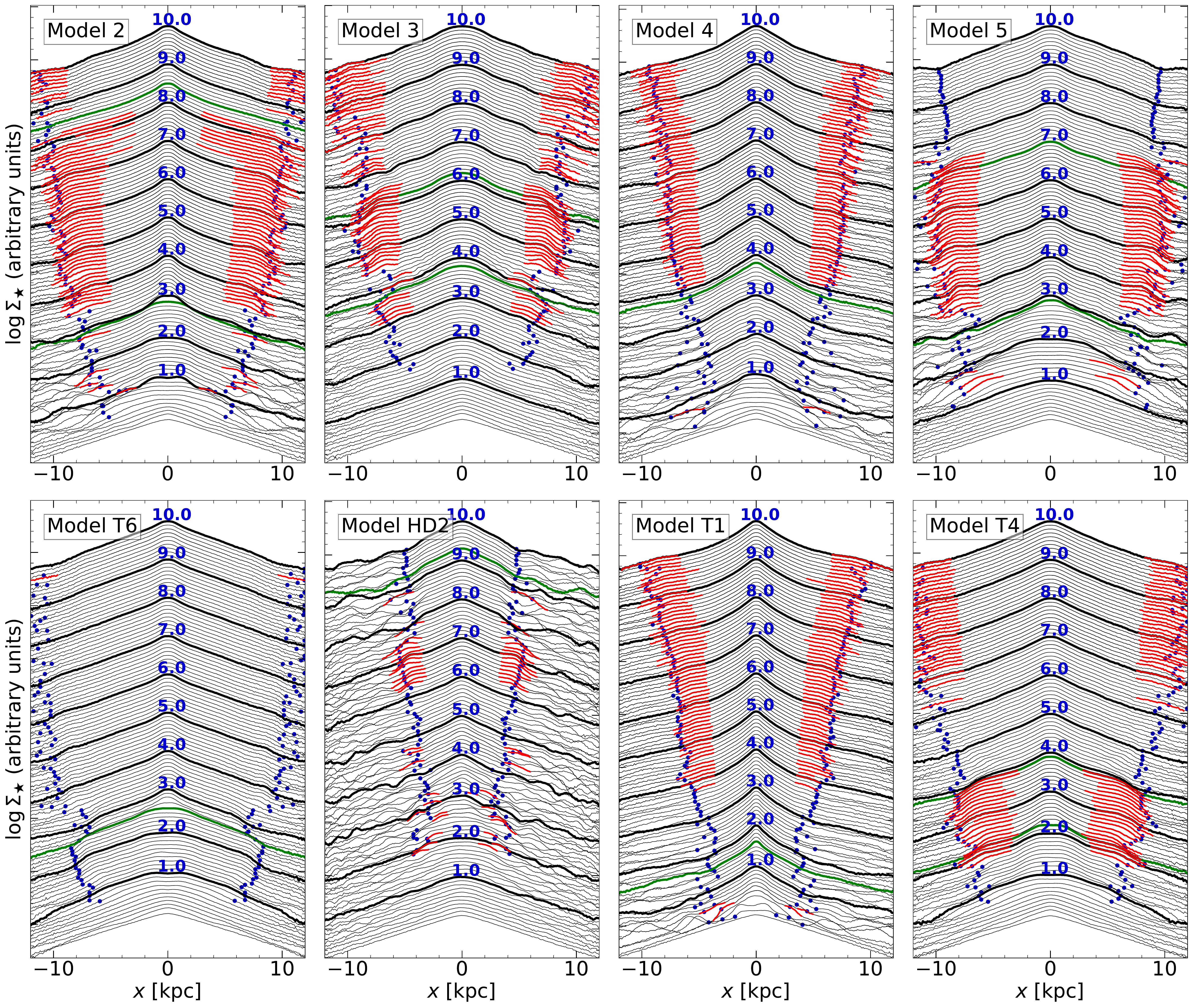}
  \caption{Evolution of the logarithmic surface density profile in the major axis cut for group B models (except model 2S which is shown in Fig.~\ref{fig:NB_Models_Evolution}). Time advances along the vertical axis. Every $1\Gyr$, the profile is highlighted in bold and the time in Gyr indicated above the maximum density. The blue dots represent the bar radius. If the \textsc{SRA} recognizes a shoulder, its extent (inner to outer edge) is shown in red. The times of major buckling episode peaks are shown in green. Shoulders, if present, usually appear after first buckling.}
  \label{fig:B_Models_Evolution}
\end{figure*}

\begin{figure*}
  \centering
  \includegraphics[width=\hsize]{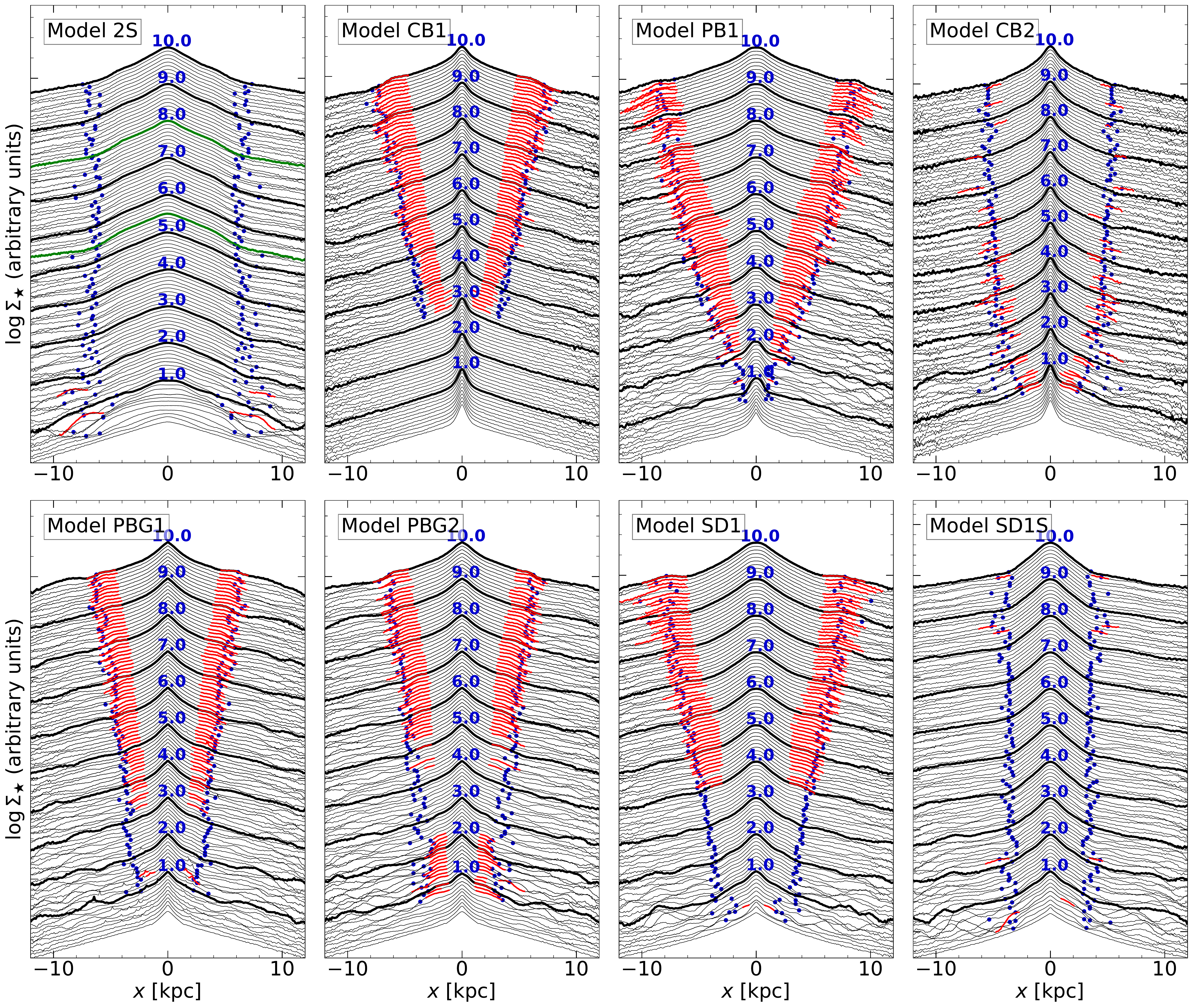}
  \caption{Evolution of the logarithmic surface density profile in the major axis cut for group WNB models, and buckling model 2S (top left). Time advances along the vertical axis. Every $1\Gyr$, the profile is highlighted in bold and the time in Gyr indicated above the maximum density. The blue dots represent the bar radius. If the \textsc{SRA} recognizes a shoulder, its extent (inner to outer edge) is shown in red. The times of the major buckling peak in model 2S are shown in green. Shoulders, if present, grow steadily soon after bar formation.} 
  \label{fig:NB_Models_Evolution}
\end{figure*}

To ensure consistency of treatment, we apply the \textsc{SRA} in an automated fashion to each time step in all models.
We apply it to nine group B (buckling) models, and seven group WNB (weak or non-buckling) models.
Figs.~\ref{fig:B_Models_Evolution} and \ref{fig:NB_Models_Evolution} show the evolution of the major axis density profiles for groups B (less model 2S) and WNB (plus model 2S), respectively. In these plots, time increases vertically with regularly spaced markers in Gyr. The blue symbols show the bar radial extent. Shoulders, when recognized by the \textsc{SRA}, are marked in red. For group B models, major buckling episodes (maxima in $A_{\mathrm{buck}}$) are shown by thick green lines. 

Although the \textsc{SRA} removes human subjectivity by running automatically, it is still subject to noise and transient effects. For example, transient spirals or density perturbations within a growing bar can be mistaken by the algorithm for shoulders. Often these features are seen around the time of bar formation, do not persist for more than 1--3 time steps and are highly asymmetrical about $x=0$ (\eg{} model 2 at $\sim1.5$ Gyr). We define \emph{persistent shoulders} as those which survive more than 3 consecutive time steps, and transients are those surviving for 3 or fewer consecutive time steps. We none the less retain all shoulder recognitions made by the algorithm, regardless of physical origin, since transients would be observationally indistinguishable from persistent shoulders. We disregard them in some analyses below, as stated in the text.

Including transients, this results in a dataset of 364 (of 909 or 40\%) and 375 (of 707 or 53\%) time steps with shoulders for groups B and WNB, respectively.

\subsubsection{Buckling versus weakly or non-buckling models}

The most striking difference between the two groups is the smooth growth of shoulders relatively soon after the bar's formation in most WNB models, whereas in most B models, persistent shoulders form during or after the first buckling episode (note that early spiral interference gives rise to transient shoulders in a few brief intervals) and tend to evolve more irregularly. 

The weakening of the bar at buckling in group B models is seen in their temporary retreat \citep{debattista+06, martinez-valpuesta_shlosman04}, although no bar is destroyed by buckling. In some WNB models (\eg{} PBG2 at $\sim2.6 \Gyr$), the shoulders weaken temporarily, and are not recognized by the \textsc{SRA} until they reappear later.

In Table~\ref{tab:models-shoulders-overview}, column $t_{\mathrm{sh}} - t_{\mathrm{bar}}$ shows the delay between bar formation and formation of persistent shoulders. The delays are somewhat smaller for group WNB. The exceptions are models PBG1 and SD1 (persistent shoulders form $\sim3\Gyr$ after bar formation in each case).

Column $t_{\mathrm{sh}} - t_{\mathrm{buck}}$ shows the time delay between first buckling and first detection of persistent shoulders in group B models, disregarding transients. A wide range of delays is seen, from almost immediate (models 5 \& T4) to $\sim3\Gyr$ (model 3).

\subsubsection{Relation with BP bulges}

We have verified, by inspecting the surface density in the $(x,z)$-plane, that all WNB models gradually form BP bulges. \citet{quillen02} demonstrated that this is possible by resonant trapping of stars, rather than by the more violent buckling instability, and \citet{petersen+2014} and \citet{sellwood_gerhard_2020} demonstrated this using \textit{N}-body simulations. This `slow mode' of BP growth results from the high central mass concentration in group WNB models (which have discy pseudobulges), which suppresses the buckling instability \citep{sotnikova_rodionov_2005, petersen+2014}. 

We checked for the presence of BP bulges in all models by examining height profiles, the fourth order Gauss-Hermite moments of the $z$ axis line of sight velocity and height distributions \citep{debattista+05}, and by visual inspection. Persistent shoulders in most B models develop during or after the emergence of a BP bulge. In most WNB models shoulders appear before BPs. There is a wide range of time differences between when shoulders and BPs emerge -- the two phenomena do not always appear together.

\subsubsection{Connection between shoulders and bar growth}
\label{ss:connection_shoulders_bar_growth}

\begin{figure*}
  \centering
  \includegraphics[width=\hsize]{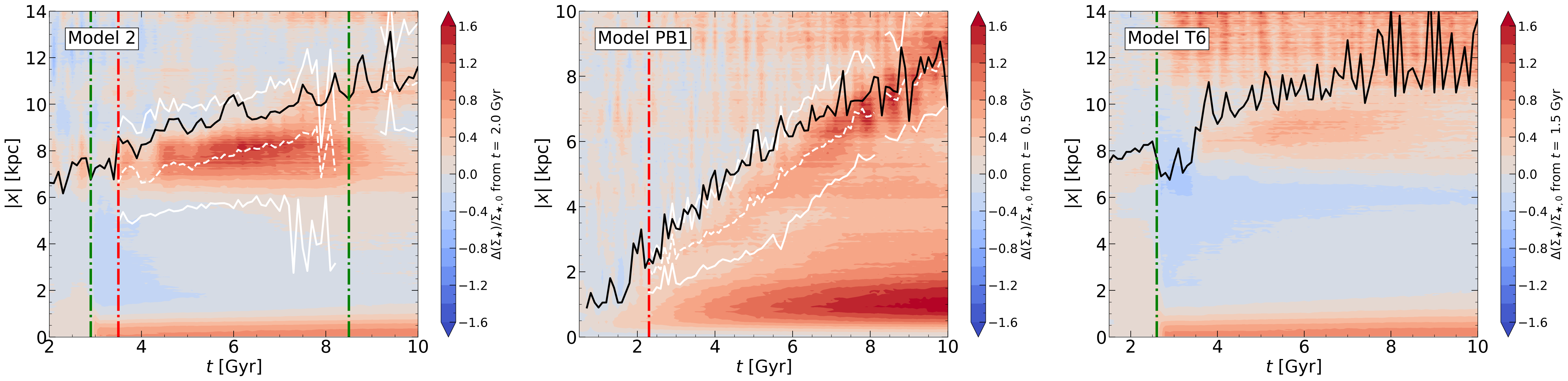}
  \caption{The fractional change in major axis surface density ($|y|\leq 1 \kpc$) for group B model 2 (left panel), group WNB model PB1 (middle panel), and model T6 whose profile remains (close to) exponential throughout its evolution (right panel). The black line is the bar radial extent. The two white lines represent the inner and outer edges of the shoulders, and the white dashed line represents the outer edge of the clavicle. Green vertical dot dash lines represent buckling times. Red vertical dot dash lines represent times when the \textsc{SRA} first recognizes persistent shoulders. Note the different scales on the $y$ axes. As the bar grows in models 2 and PB1, the density within the shoulder area increases, with the peak increase at the clavicle.}
  \label{fig:Maj_axis_heatmap.png}
\end{figure*}

\begin{figure}
  \centering
  \includegraphics[width=\hsize]{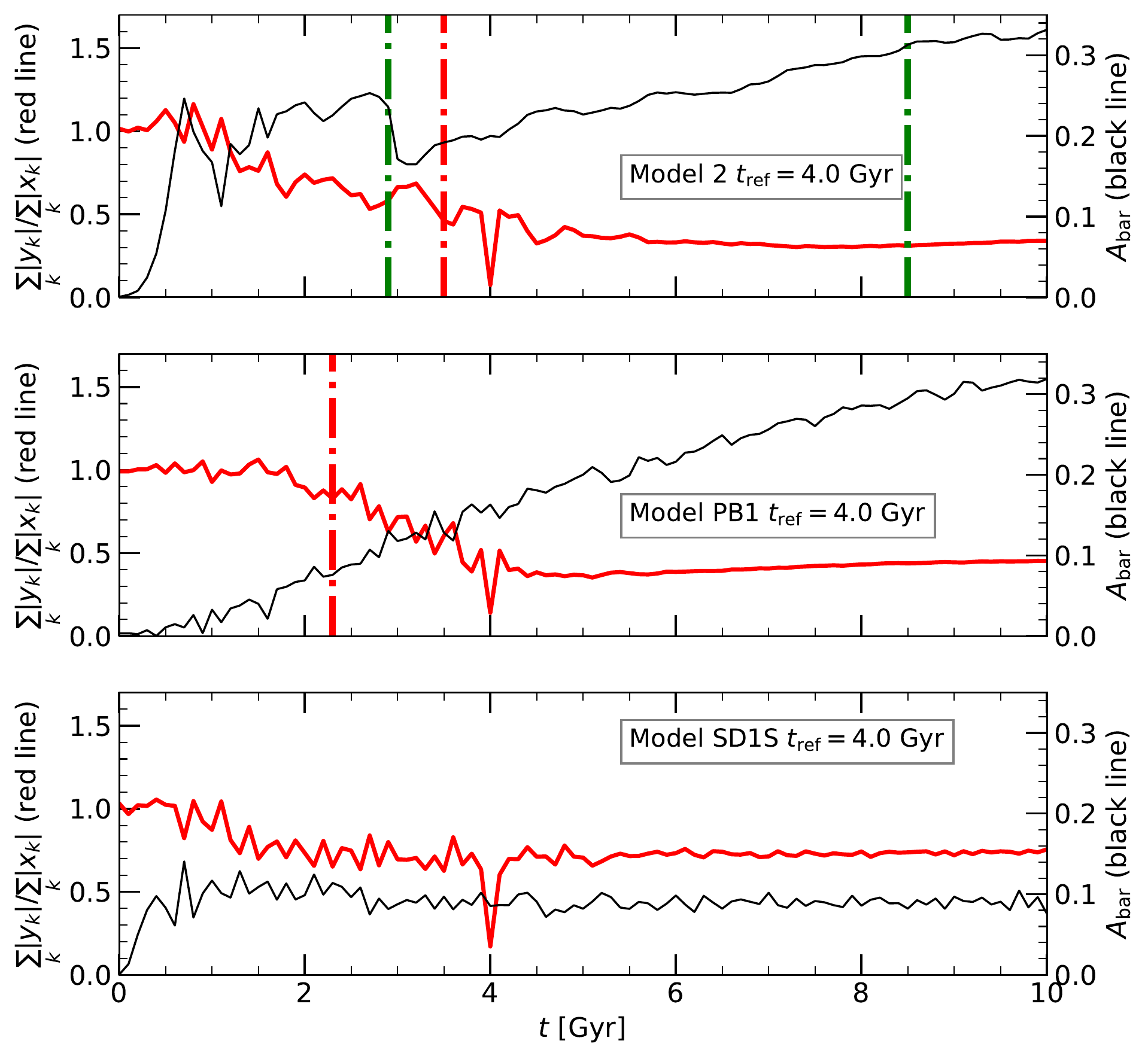}
  \caption{Evolution of the ratio $\sum_{k}{|y_k|/\sum_{k}|x_k|}$ (sum over particles) of the density distribution (red line) for three models, for particles in the shoulder at times $t_{\rm ref}$ as indicated in the annotation in each panel. For model SD1S, which lacks shoulders, we take particles in the `equivalent' shoulder region $0.66\leq|x|/R_{\mathrm{bar}}\leq1.1$ (the mean shoulder region for all shoulders in all models). Since we select particles in the shoulder regions at $t=t_{\rm ref}$, we expect and see a large drop in this ratio at that time (at $t_{\rm ref}$, the particles are in two rectangular volumes either side of $x=0$).   Black lines: $A_{\rm{bar}}$. Red vertical dot-dashed lines: times shoulders are first recognized. Green vertical dot-dashed lines: times of major buckling.}
  \label{fig:absy_over_absx}
\end{figure}

Fig.~\ref{fig:Maj_axis_heatmap.png} shows the evolution of the fractional change in major axis surface density from a time shortly before the shoulders form for two models (group B model 2 and group WNB model PB1), and model T6 whose profile remains nearly exponential (no shoulders). As the bar grows in models 2 and PB1, the density within the shoulder area increases, with the peak increase at the location of the clavicle. This suggests that additional material trapped by the growing bar `concentrates' at the clavicle. The exponential bar in T6 also shows a density increase near its end as the bar recovers from its buckling, but not enough for the model to manifest a shoulder profile. At $t\sim 8\Gyr$, bar growth falters (black line) and the (small) overdensity then reduces. These results hint at a link between growth of the shoulders and growth of the bar.

The two models with spinning prograde haloes (2S and SD1S) do not manifest any persistent shoulders. In these two models the bar growth is clearly suppressed -- the blue markers indicating $R_{\mathrm{bar}}$ hardly move in radius at all during 10 Gyr of evolution (Fig.~\ref{fig:NB_Models_Evolution}). The bar cannot grow significantly owing to the inability of a maximally spinning halo to absorb its angular momentum (although the outer disc may still absorb a small fraction) -- see \cite{athanassoula02}. Clearly then, if the bar forms but cannot grow, it will not develop shoulders, although lack of shoulders does not necessarily signify a bar which is not growing (\eg{} model T6, Fig.~\ref{fig:B_Models_Evolution}).

\subsubsection{Shoulder particle trapping by the bar}
\label{ss:shoulder_particle_trapping}

Bars grow by trapping stellar orbits. To further investigate the connection between bar growth and the shoulders, we explore how the particle orbits that come to make up the shoulder evolve with time. We do this by identifying particles present in the shoulder region at a `reference' time $t_{\rm ref}$ and computing the observed mean elongation of the distribution in $x$ and $y$ at times both before and after this time (for the same particles). At very early times, we expect these particles to lie outside the bar and thus have roughly circular distributions; as they become trapped by the growing bar, their mean distribution should become more elongated. We do this for models 5 and SD1, with reference time $t_{\rm ref} = 4.0$ Gyr; the mean elongation is computed as $\Sigma_{k} |y_{k}| / \Sigma_{k} |x_{k}|$ for all particles $k$ within the shoulders at $t_{\rm ref}$. Values of $\Sigma_{k} |y_{k}| / \Sigma_{k} |x_{k}| \sim 1$ indicate near-circular orbits. For comparison, we also perform this analysis for model SD1S, which remains exponential and does \textit{not} form shoulders; we define an `equivalent' region corresponding to the mean shoulder extent (relative to the bar size) from the models that do form shoulders.

Fig.~\ref{fig:absy_over_absx} shows a gradual reduction in $\Sigma_{k} |y_{k}| / \Sigma_{k} |x_{k}|$ from $\sim 1$, followed by an approximately constant value of $\sim 0.4$ for models 2 and PB1 (red lines). This is an indicator of trapping by the bar as orbits become elongated, and the particles in the shoulder at $t_{\rm ref}$ are largely trapped by the time the ratio reaches $\sim 0.4$. The particles in the exponential bar in model SD1S show similar trapping behaviour; however, the final (post trapping) ratio in this case is significantly higher ($\sim 0.7$), a much lower level of elongation. This indicates that these particles are not trapped into the same orbit morphology as those particles destined to be in a shoulder. The bar in this model does not grow, and no shoulders form. High elongation orbits are adopted by newly trapped particles once the bar begins its growth (we explore the orbits supporting the shoulders in Section~\ref{s:orbits}); in other words, bars are not formed with shoulders but acquire them as they grow.

\subsubsection{Pure star-forming model}
\label{ss:star_forming_model}

\begin{figure}
\centering
\includegraphics[width=\hsize]{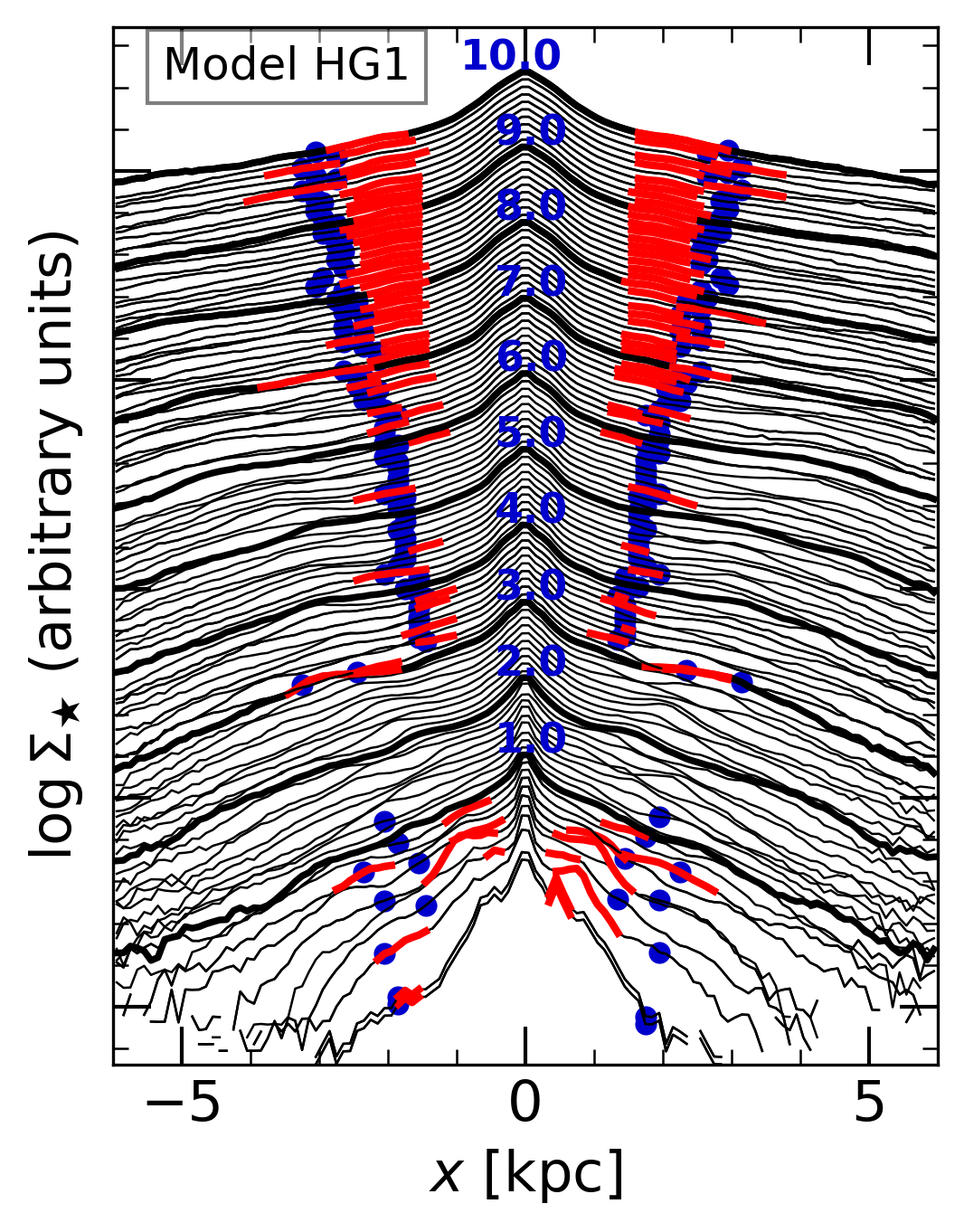}
\caption{Evolution of the logarithmic surface density profile in the major axis cut for the pure star-forming model HG1. Time advances along the vertical axis. Every $1\Gyr$, the profile is highlighted in bold and the time in Gyr indicated above the maximum density. The blue dots represent the bar radius. If the \textsc{SRA} recognizes a shoulder, its extent (inner to outer edge) is shown in red. The pure star-forming model forms persistent shoulders as do the $N$-body models.}
\label{fig:708_Shoulder_Evolution_1}
\end{figure}

All models considered so far had stellar discs as part of the initial setup. To determine if our conclusions are strictly a result of these initial conditions, we examine model HG1 (see section~\ref{s:star_forming_models}). This model does not buckle strongly. As shown by \citet{athanassoula_2013}, in the presence of significant amounts of gas, bars are expected to form later and be weaker than in gas-free models. Model HG1 shows these characteristics; it forms a smaller bar at $\sim 3$ Gyr. From that point the bar grows moderately but steadily ($R_{\mathrm{bar}}\sim 3$ kpc at $t=10$ Gyr). The bar is weaker than in models with initial discs, having max $A_{\mathrm{bar}}\sim0.12$. Fig.~\ref{fig:708_Shoulder_Evolution_1} shows the shoulder evolution for this model. Persistent shoulders only emerge once the bar grows significantly in radius at $\sim6$ Gyr, consistent with our earlier analysis. We have verified, by examining the height profile, fourth-order Gauss-Hermite moments of the $(x,y)$ line of sight velocity and height distributions \citep{debattista+05} along the major axis that a BP bulge is present when the shoulders appear. Since these results are consistent with those for the models with initial discs considered above, we conclude that those initial conditions do not lead to artificial shoulder formation. None the less, the bars in the pure $N$-body simulations grow rapidly and quickly become larger (mean $R_{\rm bar}=8.4$ kpc at $t=10\Gyr$) than those observed \citep{erwin_2019}.

\subsection{Quantitative properties of the shoulders}
\label{ss:sh_quant_properties}

We have shown that persistent shoulders develop as a part of bar growth, and are accompanied by BP bulges. We now examine quantitatively the relationship between bar/BP growth and shoulder evolution.

\subsubsection{Shoulder outer edge location} 
\label{ss:outer_edge_location}

For all models which have shoulders at $t=10\Gyr$, we compute the location of the shoulder edge ($R_{\mathrm{sh}}$), normalised to $R_{\mathrm{bar}}$ (bar radius) and $R_{\mathrm{BP}}$ (BP bulge `radius'). Groups B and WNB are consistent within the errors. On average, $R_{\mathrm{sh}}$ lies just beyond the bar radius, with $R_{\mathrm{sh}}/R_{\mathrm{bar}} = 1.16\pm0.06$. There is little evolution of this ratio in most models. With respect to the BP bulge radial extent, we find $R_{\mathrm{sh}}/R_{\mathrm{BP}} = 2.30\pm0.27$. This is consistent within errors with \citet{lutticke+00}, who found a ratio of $2.7\pm0.3$ in a sample of 43 edge-on barred galaxies.

\subsubsection{Shoulder edge, strength and flatness}

\begin{figure*}
  \centering
  \includegraphics[width=\hsize]{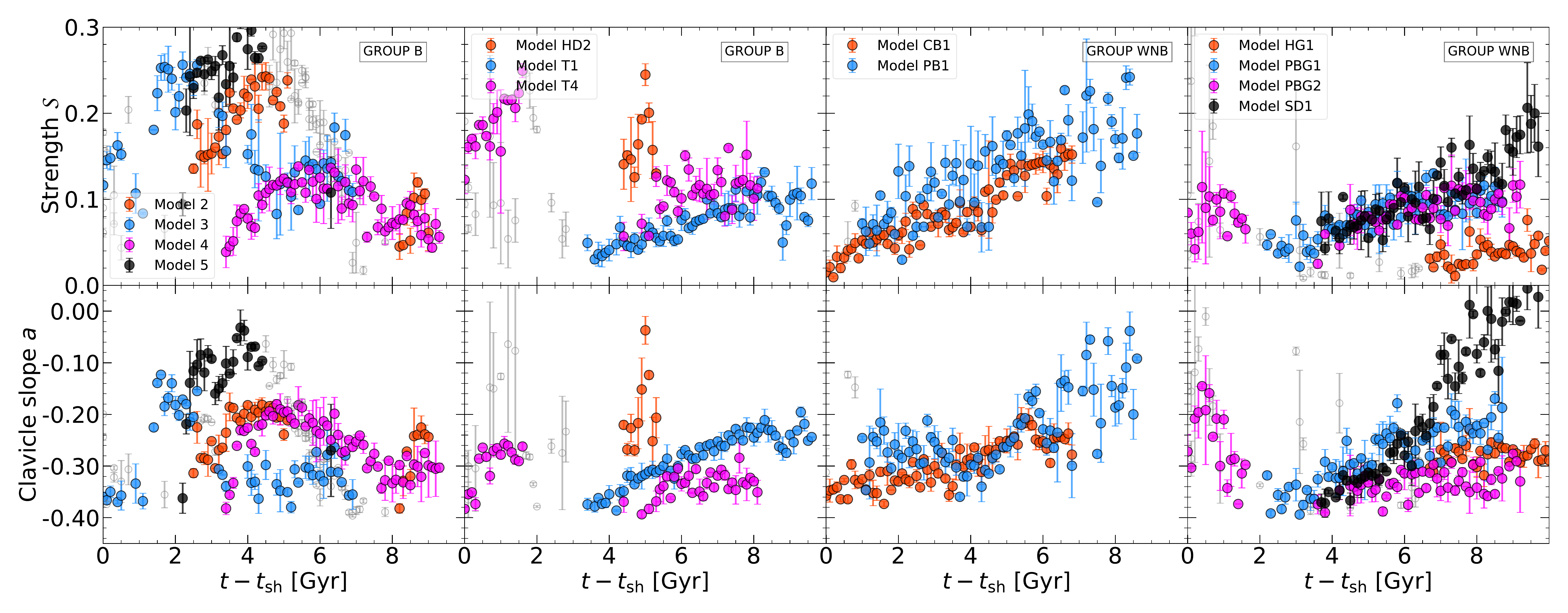}
  \caption{The evolution of shoulder edge, $R_{\mathrm{sh},}$ normalised to $R_{\mathrm{bar}}$, strength, $\mathcal{S}$, and slope, $a$, as a function of time since shoulder formation. Transient shoulders, and periods of decline during secondary buckling, are shown with grey symbols to focus on periods of steady shoulder growth.}
  \label{fig:Shoulder_param_evol}
\end{figure*}

Fig.~\ref{fig:Shoulder_param_evol} shows the evolution of the shoulder strength $\mathcal{S}$ and slope $a$ (\ie{} the flatness), starting at the time of first shoulder detection for each model.

Usually, $\mathcal{S}$ increases with time in both groups (unless secondary buckling or periods of shoulder weakening occur, \eg{} model 5 at $\sim5 \Gyr$ after first shoulder recognition), and the shoulder evolves towards 0 slope (\ie{} flatter). As the shoulders grow they contain an ever higher fraction of excess mass. We have verified that in most models, $R_{\mathrm{sh}}$, when normalised by $R_{\mathrm{bar}}$, remains within a relatively narrow range, supporting the idea that shoulders develop as part of the bar's evolution. This is consistent with the simulations of \citet{bureau_athanassoula05}, who found that the extent of the `plateaux' grew with time as the bar lengthened.

\subsubsection{Correlations with bar and BP properties} 
\label{sh_correlations}

\begin{figure*}
  \centering
  \includegraphics[width=\hsize]{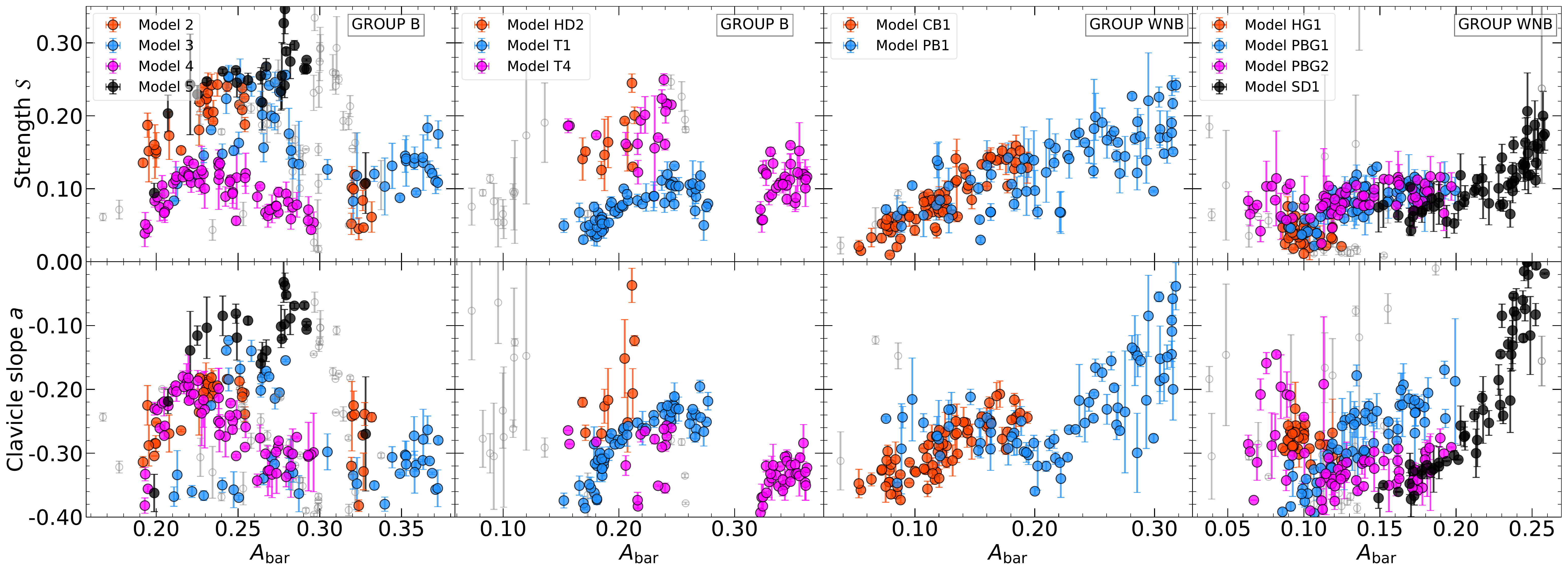}
  \caption{Plots of shoulder strength and slope with bar strength $A_{\mathrm{bar}}$. Transient shoulders, and periods of decline during secondary buckling, are shown with grey symbols to focus on periods of steady shoulder growth.}
  \label{fig:Bar_correls}
\end{figure*}

Fig.~\ref{fig:Bar_correls}  plots $\mathcal{S}$ and $a$ versus $A_{\mathrm{bar}}$. There is no general correlation (see also Fig.~\ref{fig:absy_over_absx} where shoulders form at different bar strengths); however, for around half of the individual models -- many WNB models and models T1, 2 and 4 before shoulder weakening -- the strengths of the shoulders and bar are correlated, albeit with significant scatter (top row), with a similar, but weaker relation between $a$ and $A_{\mathrm{bar}}$ (bottom row). $\mathcal{S}$ and $a$ are similarly correlated with $R_\mathrm{{bar}}$ (not shown). For these models, the stronger and longer the bar, the stronger and flatter the shoulder. We note that the slope for model SD1 becomes positive towards the end of the simulation, \ie{} up-bending shoulders (see Fig.~\ref{fig:NB_Models_Evolution}).

\begin{figure}
  \centering
  \includegraphics[width=\hsize]{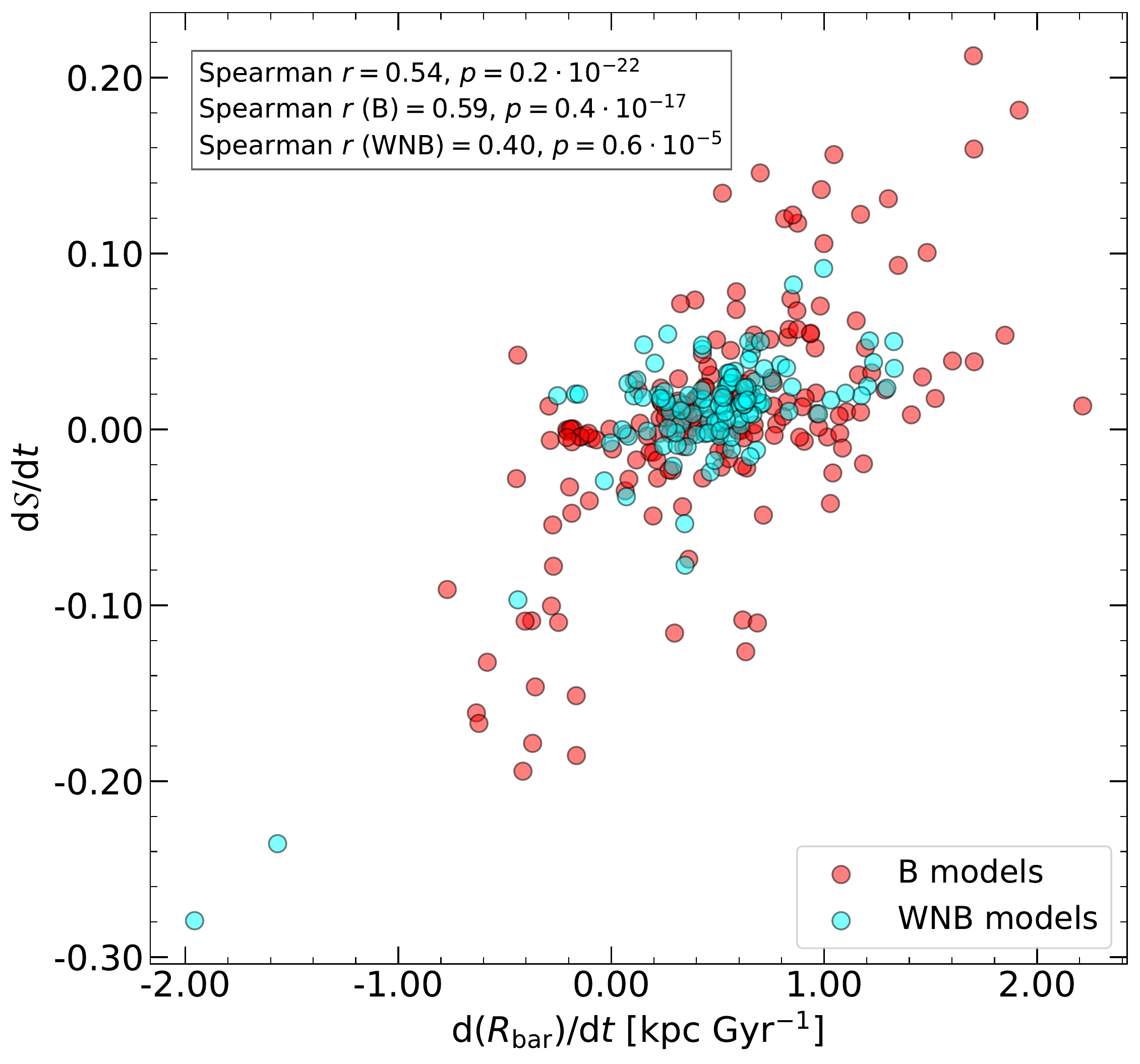}
  \caption{Rate of growth of the shoulder strength (excess mass fraction), d$\mathcal{S}/$d$t$ versus rate of growth of the bar radius, d$R_\mathrm{bar}/$d$t$, for all models. Red points are B models, cyan points are WNB models. Transients are excluded. To reduce noise, we average over four time steps. Spearman rank correlation coefficients and associated $p$-values are shown in the annotation.}
  \label{fig:drbar_dt}
\end{figure}

In Fig.~\ref{fig:drbar_dt}, we show the relationship between d$\mathcal{S}/$d$t$ and d$R_\mathrm{bar}/$d$t$ for all models, split between B and WNB models. We disregard transients and smooth out some of the noise by averaging every four time steps. Note that we have few points with d$R_\mathrm{bar}/$d$t < 0$, since there are relatively few time steps where the bar is shrinking in radius. Although still somewhat noisy, the plot shows that as the rate of increase in bar radius rises, so does the rate at which excess mass is trapped in the shoulder. This, coupled with the demonstration above that the longer the bar the stronger the shoulders, is evidence of a link between growth of the bar, and strength of the shoulders.

\begin{figure*}
  \centering
  \includegraphics[width=\hsize]{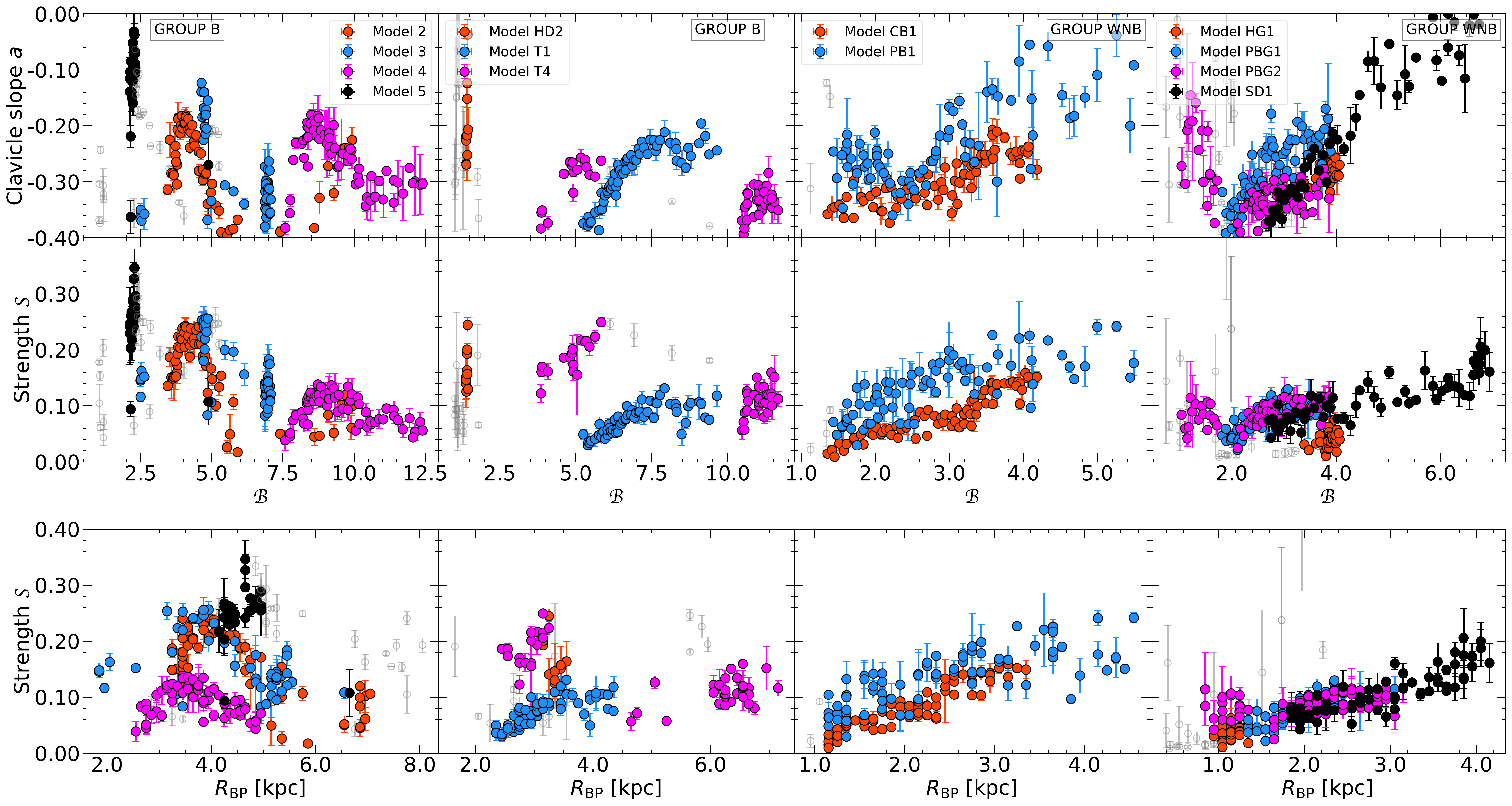}
  \caption{Plots of shoulder strength and slope with BP bulge strength, $\mathcal{B}$, and radial extent, $R_{\mathrm{BP}}$. Transient shoulders, and periods of decline during secondary buckling, are shown with grey symbols to focus on periods of steady shoulder growth.}
  \label{fig:BP_correls}
\end{figure*}

Fig.~\ref{fig:BP_correls} plots shoulder parameters versus BP size and strength $\mathcal{B}$. Within most WNB models, the shoulders become flatter and stronger as the BP bulge becomes stronger. For most buckling models, the shoulders grow alongside a BP bulge whose strength remains relatively constant. For models exhibiting the correlation between $R_{\mathrm{BP}}$ and $\mathcal{B}$, a variety of slopes is observed.

\section{Shoulder dissolution}
\label{s:dissolution}

In most models shoulders persist once established (Figs.~\ref{fig:B_Models_Evolution}, ~\ref{fig:NB_Models_Evolution}). In some models, however they dissolve with little impact on the bar strength (models 5 and T4 in Fig.~\ref{fig:B_Models_Evolution}). For model 5, the shoulders exist only between the first and second buckling. The second buckling is responsible for shoulder dissolution. Dissolution does not need to be permanent: model 3 suffers two buckling episodes, but its shoulders recover soon after the second.

\begin{figure}
  \centering
  \includegraphics[width=\hsize]{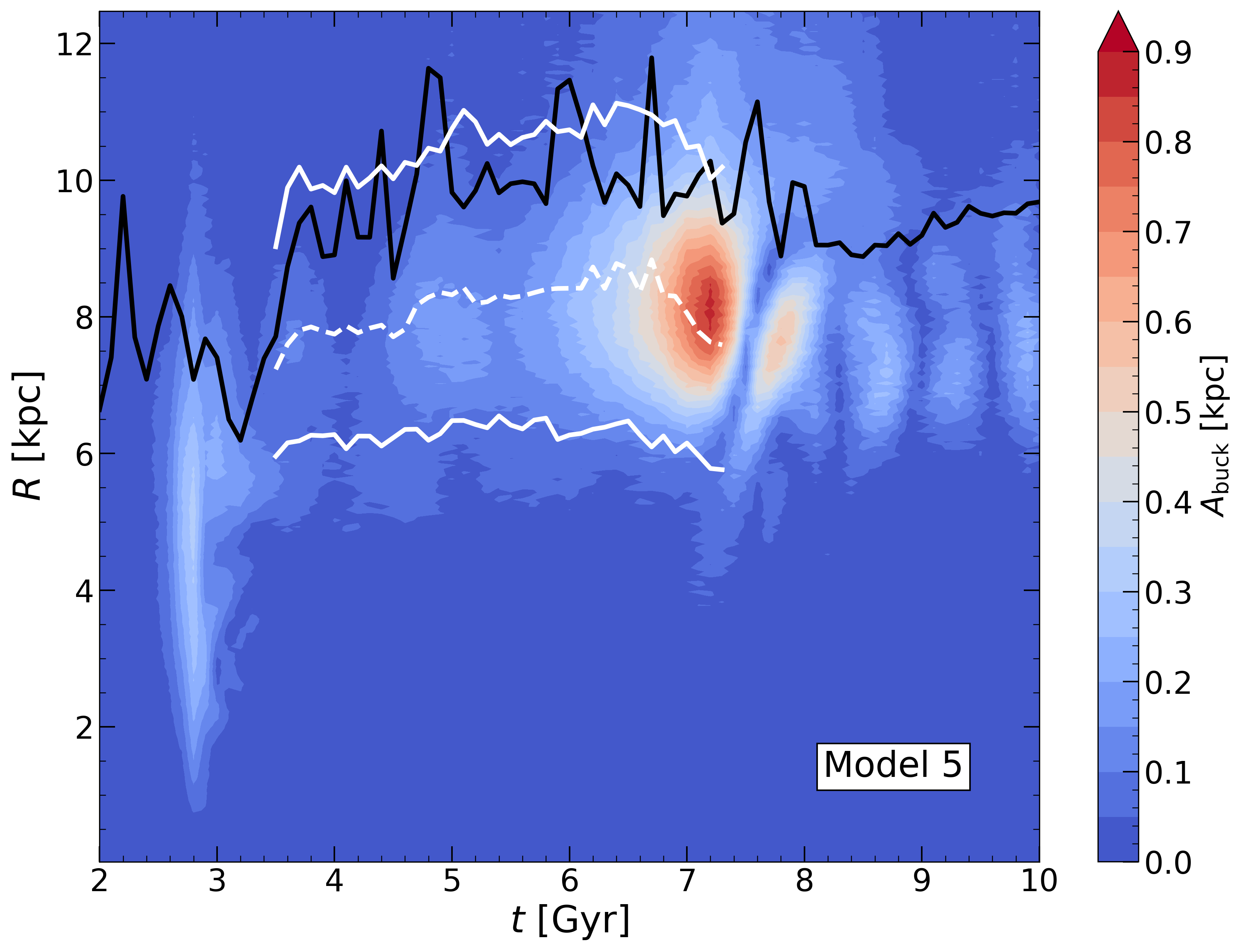}
  \caption{Evolution of $A_{\mathrm{buck}}$ (by cylindrical $R$) for model 5 from $t=2 \Gyr$. The black line represents the bar radius, the white dashed line represents the outer edge of the clavicle and the two solid white lines represent the location of the inner and outer edges of the shoulder. A second buckling tends to occur in the shoulder region and destroys the shoulder.}
  \label{fig:D7_radial_A_buck_heatmap}
\end{figure}

Fig.~\ref{fig:D7_radial_A_buck_heatmap} shows the radial evolution of $A_{\mathrm{buck}}$ for model 5. The first buckling episode ($t\sim2.8 \Gyr$) results in a BP bulge. Persistent shoulders form at $t\sim3.5 \Gyr$ and grow steadily. The second buckling at $t\sim6.8\Gyr$ occurs in the shoulder area, with $A_{\mathrm{buck}}$ peaking between the middle of the clavicle to the end of the bar although there is little impact on the bar strength (Fig.~\ref{fig:barbuck_amp}). The portion of the disc at radii smaller than the inner clavicle edge is not affected by the second buckling (consistent with the simulation of \citet{lokas_2019}, who found a second buckling was concentrated in the outer part of the bar, and did not impact the bar strength). As the second buckling begins, the entire shoulder retreats somewhat before vanishing. This happens rapidly ($\sim0.7 \Gyr$ separate the peak in $A_{\mathrm{buck}}$ at second buckling and the time when the \textsc{SRA} no longer recognizes shoulders). Qualitatively similar results are observed for models 2, 3 and T4, except that in model 3, the shoulders regrow almost immediately. Major secondary buckling is thus focused in the shoulder area, and rapidly turns a flat bar profile into an exponential one. In some cases shoulders are able to regrow quickly. This is dependent on the bar's ability to capture sufficient additional material after the second buckling episode. We have verified that the corotation radius ($R_\mathrm{CR}$) increases gradually in all models except those with spun-up halos (D6S and SD1S). It might be that, after a second buckling, a relative dearth of material beyond $R_\mathrm{CR}$ prevents quick reformation of the shoulders, since the bar has less additional mass to trap than previously, when $R_\mathrm{CR}$ was lower. Models 5 and T4 may be exhibiting this behaviour after their second buckling. Shoulder dissolution therefore, may either be temporary or long lasting.

Secondary buckling is not the only way shoulders dissolve. In models HD2 at $t\sim7.5$ (Fig.~\ref{fig:B_Models_Evolution}) and PBG2 at $t\sim2.5$ Gyr (Fig.~\ref{fig:NB_Models_Evolution}), persistent shoulders dissolve without buckling. In these cases, we have verified that shoulder dissolution occurs via strong spirals perturbing the ends of the bar, disrupting its morphology and reducing its size.

Model T6 has no shoulders detected by the \textsc{SRA}, before or after buckling, in contrast with model T4, which has a similar bar radius evolution. While close inspection of model T6 reveals hints of shoulders near the end of the bar at $\sim3$ Gyr, this is below our chosen detection threshold ($T\sim0.5$ versus the threshold $0.4$, so they are very weak), persisting until $\sim5.5$ Gyr. The difference between models T4 and T6 appears to be caused by the thicker disc of T6 relative to T4 (median $|z|=0.26$ versus $0.10$ kpc at $t=0$, respectively). As such, there is more material at large heights above the bar, which dilutes any shoulders that would form from recent trapping.

\section{Observational considerations}
\label{s:implications_for_observations}

\subsection{Shoulders in projection}

\begin{figure*}
  \centering
  \includegraphics[width=\hsize]{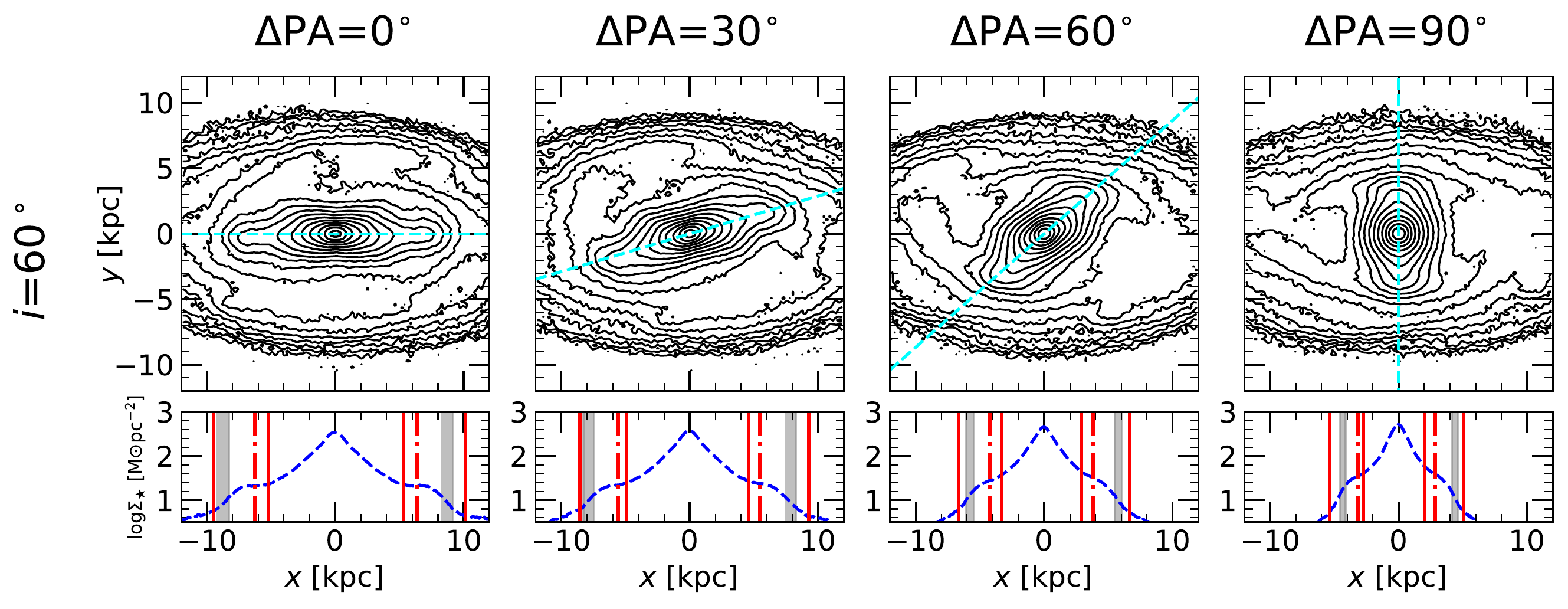}
  \caption{Surface density contours in the projected $(x,y)$-plane for model 2 at $t=5 \Gyr$. The model has been rotated to inclination $i= 60\degrees$ and intrinsic bar position angles $\Delta$PA $= 0\degrees$, $30\degrees$, $60\degrees$ and $90\degrees$. Within each panel, the dashed cyan line represents the bar's projected major axis. Beneath each $(x,y)$ panel is the logarithmic surface density (in arbitrary units) along the bar's projected major axis (projected $|y|\leq 1 \kpc$). The grey area represents the projected bar's radial extent. In all cases, the \textsc{SRA} recognizes shoulders; the vertical dot-dashed lines represent the clavicle centres and the vertical red lines mark the inner and outer boundaries of the shoulders.}
  \label{fig:D6_t50_projection}
\end{figure*}

\begin{figure*}
  \centering
  \includegraphics[width=\hsize]{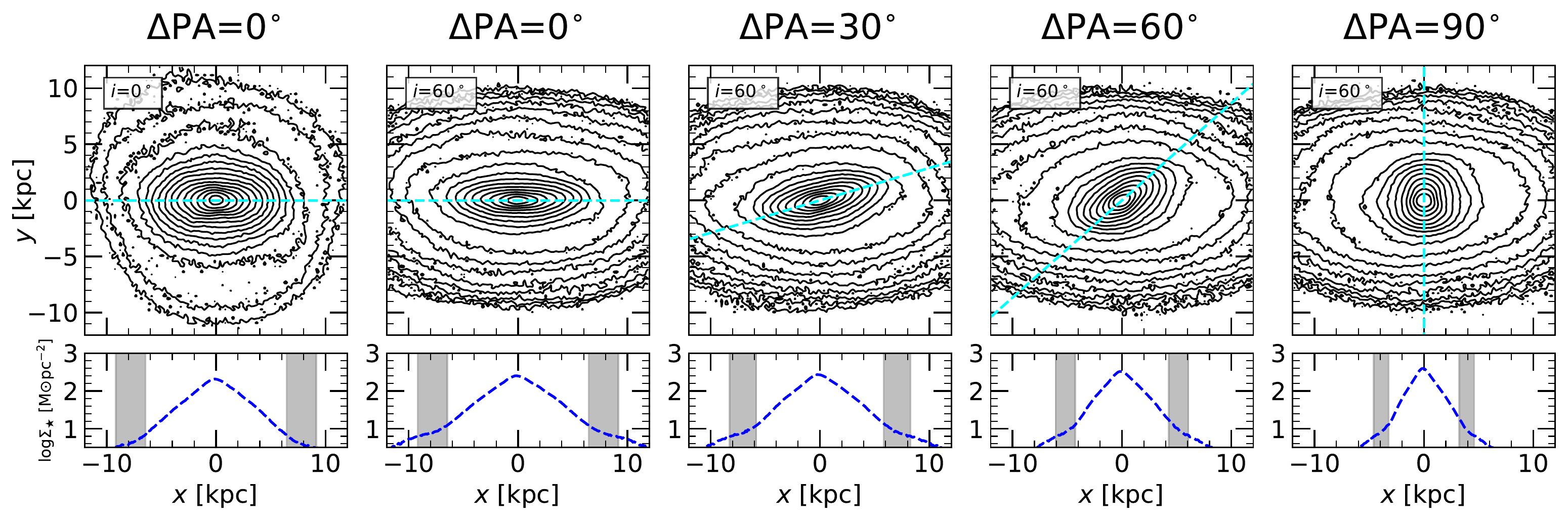}
  \caption{Surface density contours in the projected $(x,y)$-plane for model 2S at $t=5$ Gyr. The first column shows the face on projection. In the remaining columns, the model has been rotated to inclination $i=$ 60$\degrees$ and intrinsic bar position angles $\Delta$PA = 0$\degrees$, 30$\degrees$, 60$\degrees$ and 90$\degrees$. Within each panel, the dashed cyan line represents the projected bar's major axis. Beneath each ($x,y$) panel is the logarithmic surface density (in arbitrary units) along the projected major axis (projected $|y|\leq 1$ kpc). The grey area represents the projected bar radius. The \textsc{SRA} does not recognize shoulders at any bar position angle.}
  \label{fig:D6S_t50_projection}
\end{figure*}

\begin{figure}
  \centering
  \includegraphics[width=\hsize]{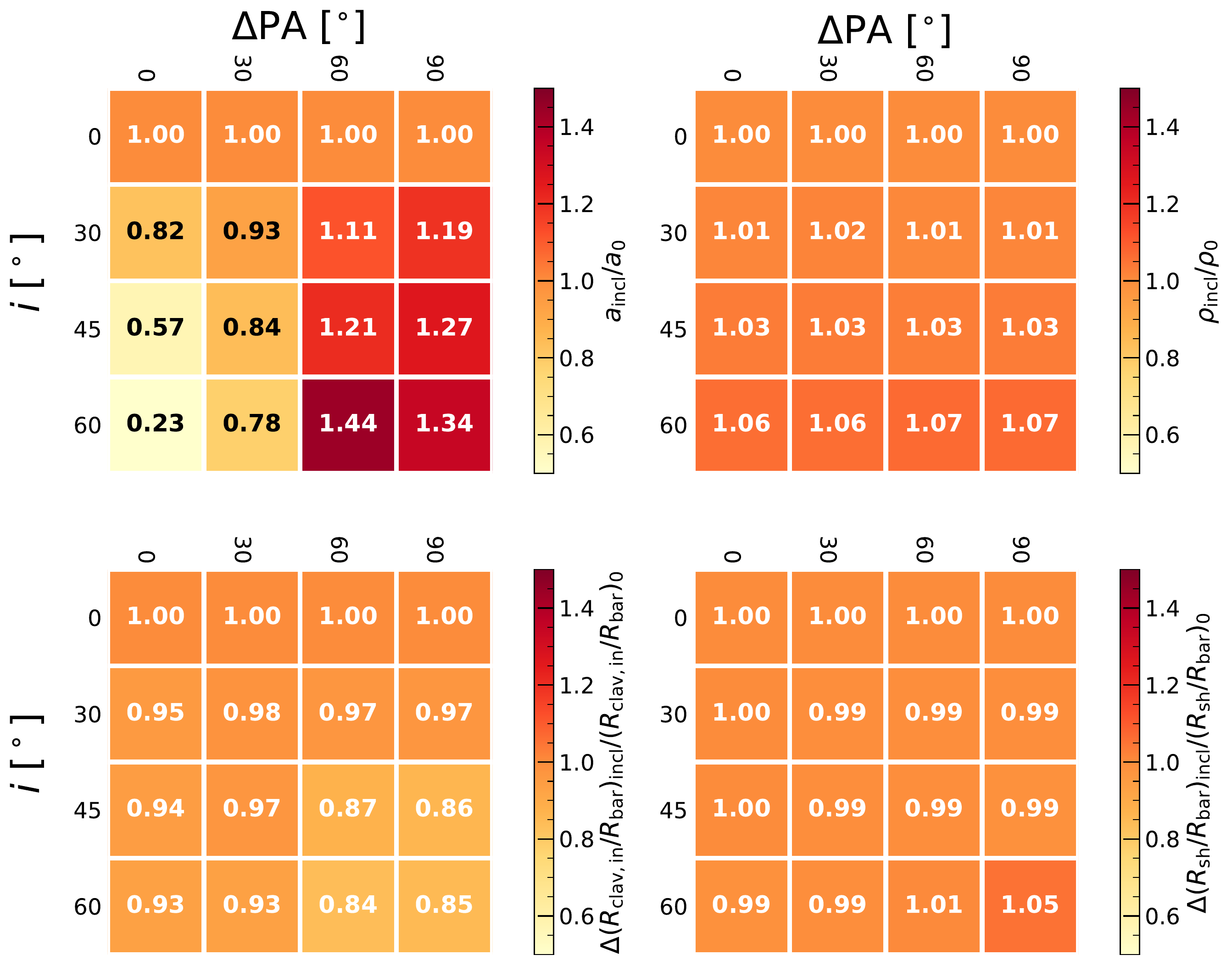}
  \caption{Inclined (subscript `incl') to face on (subscript `0') ratios for observable shoulder parameters for model 2 at $t=5 \Gyr$. The model has been rotated to different inclinations, $i$, and intrinsic bar position angles, $\Delta$PA, as indicated. Top panel, left to right: slope $a$, ratio of peak to clavicle logarithmic surface density $\rho$. Lower panel, left to right: $R_{\mathrm{clav,in}}/R_{\mathrm{bar}}$, $R_{\mathrm{sh}}/R_{\mathrm{bar}}$.}
  \label{fig:D6_t50_projection_sh_deltas}
\end{figure}

We now explore whether projection affects the recognition and observed properties of shoulders. We examine model 2 at various inclinations $i<60\degrees$ and relative bar position angles $\Delta$PA. The \textsc{SRA} recognizes shoulders for every combination of $i$ and $\Delta$PA, implying that shoulders, if they exist, would be observed at all inclinations up to $\sim60\degrees$. For example, Fig.~\ref{fig:D6_t50_projection} shows the model and shoulder detection for $i=60\degrees$ at $t=5 \Gyr$ (when the BP bulge and face-on shoulders are well established), for various bar position angles. \citet{erwin_debattista13} showed that moderately inclined galaxies with a BP bulge have isophotes with a boxy inner part from the thickened portion of the bar, accompanied by spurs. They also showed that when the bar's position angle is beyond $\sim50\degrees{}$ from the major axis, such a morphology may not be apparent. Fig.~\ref{fig:D6_t50_projection} shows that in such cases, the shoulders are still recognized. 

In contrast, model 2S with a spinning halo does not manifest shoulders at any time step. Fig.~\ref{fig:D6S_t50_projection} shows the model and shoulder detection for $i=60\degrees$ at $t=5 \Gyr$, with the first column showing the face-on projection. Shoulders are not recognized at any combination of $i$ and $\Delta$PA.

Fig.~\ref{fig:D6_t50_projection_sh_deltas} shows the ratios of inclined to face-on values for some of the observable shoulder parameters, when viewed in projection. Most parameters are not substantially ($\gtrsim15\%$) affected.
However, $a$ (\ie{} flatness, top left panel) \emph{is} strongly affected, and this parameter determines whether a shoulder is recognized. Values less than 1 indicate that a shoulder becomes flatter, and greater than 1 indicate that it becomes steeper, \ie{} less shoulder-like. For higher $i$ and smaller $\Delta$PA ($\lesssim 30\degrees$), shoulder profiles become flatter. For higher $\Delta$PA, shoulder profiles appear less flat -- potentially resulting in them disappearing altogether if they are intrinsically weak. Some caution is therefore needed when interpreting flat bar profiles in projection.

\subsection{Buckling bars and shoulders}
\label{ss:buckling_bars_and_shoulders}

\begin{figure}
  \centering
  \includegraphics[width=\hsize]{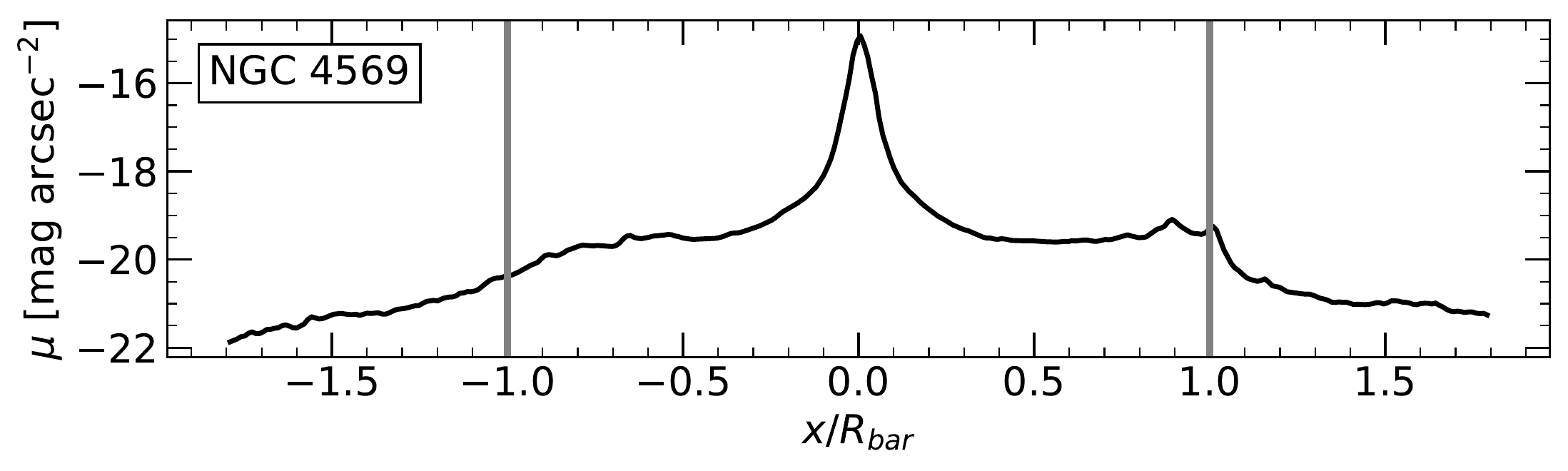}
  \caption{The major axis surface brightness profile in the \textit{Spitzer} IRAC $3.6\micron$ band for NGC~4569 \citep{kennicutt_03}, which is currently buckling. The major axis has been scaled to the bar radius $R_{\mathrm{bar}}$ and the thick vertical grey lines mark the size of the bar.}
  \label{fig:n4569}
\end{figure}

\citet{erwin_debattista16} presented the first direct detection of buckling bars, in NGC~4569 and NGC~3227 through examination of their isophotes. Using simulations, they noted that buckling galaxies are expected to exhibit spurs offset on the same side of the projected major axis, together with trapezoidal, rather than boxy, inner isophotes (their Fig.~2). Fig.~\ref{fig:n4569} shows the surface brightness profile along the bar major axis (on the sky, not deprojected) of NGC~4569 ($i=69\degrees$, $\Delta\mathrm{PA}=26\degrees$). It is a system with shoulders within the bar (albeit with significant star formation contaminating the $x>0$ profile). Is it possible to have both the isophotal morphology of buckling and shoulders?

\begin{figure}
  \centering
  \includegraphics[width=\hsize]{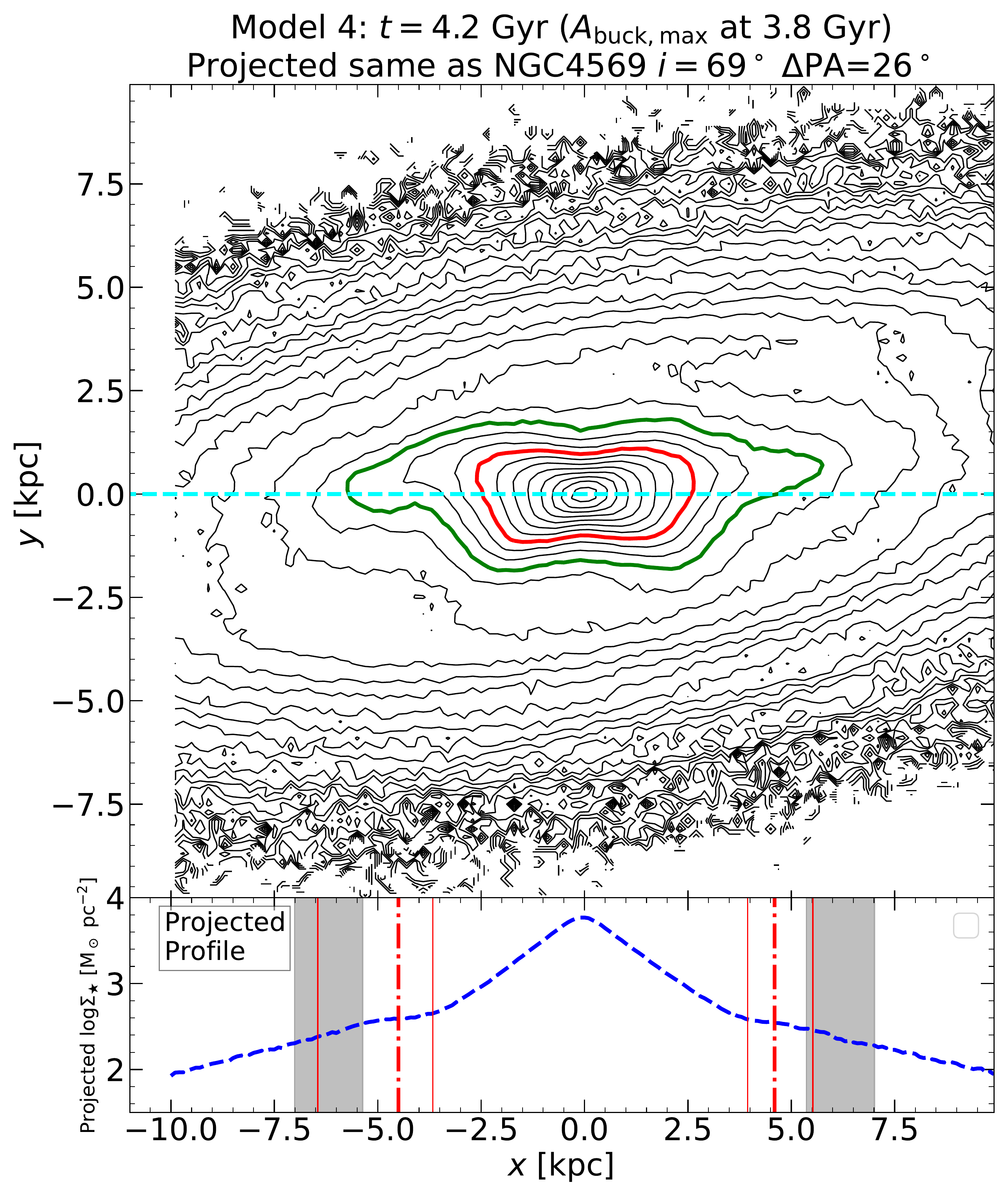}
  \caption{Upper panel: $\log\Sigma_{\bigstar}$ density contours in the projected $(x,y)$-plane for model 4 during its buckling at time $t=4.2 \Gyr$, $0.4\Gyr$ after maximum $A_{\mathrm{buck}}$. The model has been rotated to match the inclination and bar relative position angle of NGC~4569. The dashed cyan line represents the projected bar's major axis, which has been rotated to lie along the projected $x$-axis. A sample trapezoidal inner isodensity contour is shown in red, and in green is a contour with spurs offset on the same side of the major axis; these are the observational signatures of a buckling bar. Lower panel: $\log\Sigma_{\bigstar}$ along the projected bar major axis, $x$ (projected $|y|\leq 1 \kpc$). The grey area represents the projected bar radius, the vertical dot-dashed lines represent the clavicle centres as identified by the \textsc{SRA}, and the thin vertical red lines mark the boundaries of the shoulders.}
  \label{fig:D8_t42_like_NGC4569_projection}
\end{figure}

Fig.~\ref{fig:D8_t42_like_NGC4569_projection} demonstrates that it is indeed possible: we show model 4 shortly after peak buckling, at $t=4.2\Gyr$, rotated to the same orientation as NGC~4569.
The model presents the same isophotal morphology as NGC~4569, and the \textsc{SRA} recognizes shoulders in this projected profile. We find a similar signal in model 3, and in models 5 and T1 during their second bucklings (for a short time before the shoulders dissolve), but not their first. Thus, the models are consistent with those of NGC~4569, and the coexistence of shoulders and buckling does not necessarily imply a second buckling.

\section{Imprint of shoulder-supporting orbits} 
\label{s:orbits}

\begin{figure*}
  \centering
  \includegraphics[width=\hsize]{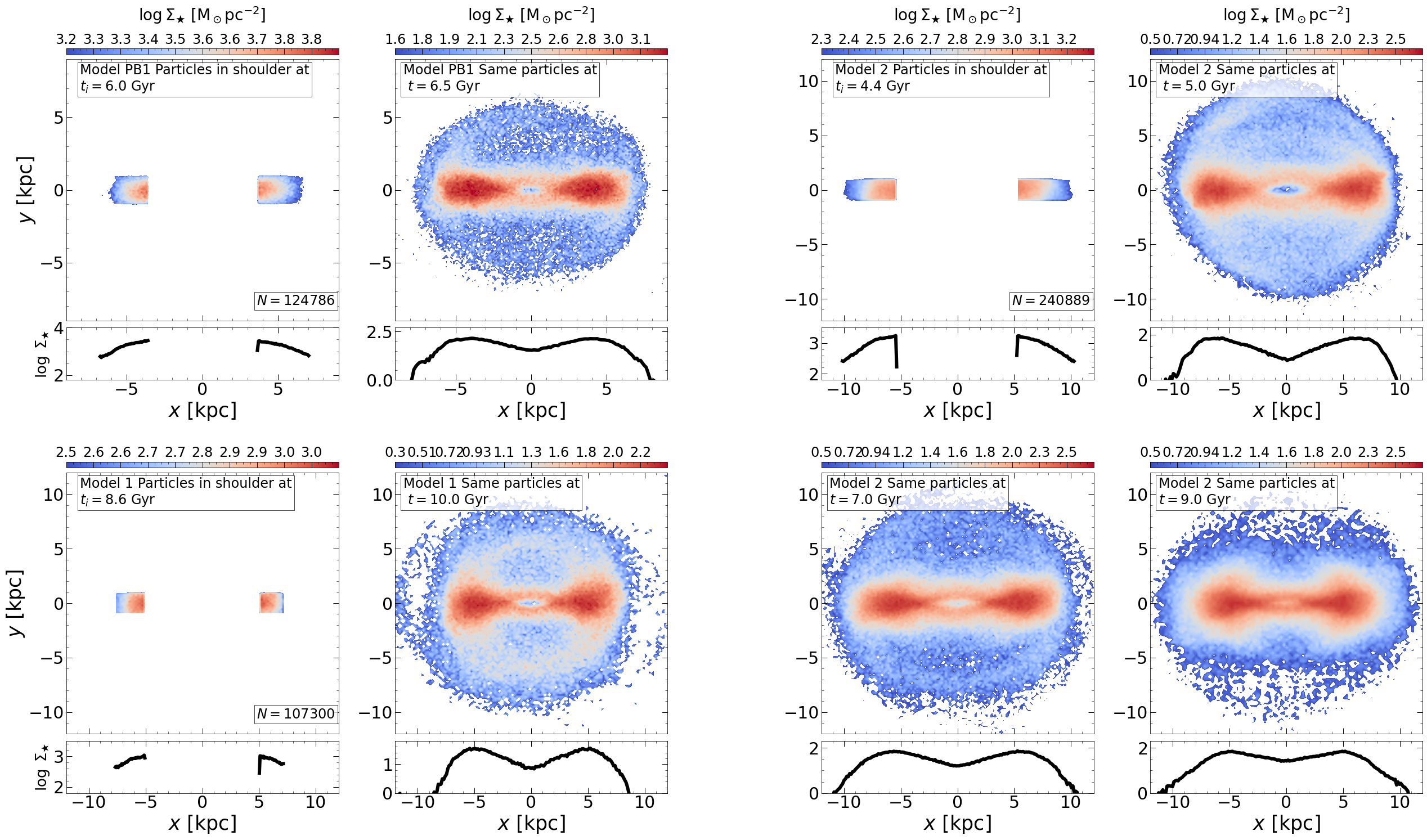}
  \caption{Surface density in the $(x,y)$-plane at times indicated in the annotations in each panel, for those particles located within the shoulders at an earlier time step $t_i$. Within each pair of plots (except the pair at lower right), the left plot shows the particles in the shoulder at $t_i$, and the number of particles is indicated in the lower right of each panel ($N$). The right plot shows the same particles later in the model's evolution. Beneath each panel we show $\log\Sigma_{\bigstar}$ for the particles along the major axis ($|y|\leq1$ kpc). Note the different scales on the plots. The pair at lower right show the particles for model 2 at $t_i=4.4$ Gyr, but at later times $7$ and $9$ Gyr. The plots show the loop-like morphology of the underlying orbits, and the lower right pair shows how the orbits of a given set of shoulder particles tend to librate in time, diluting these particles' contribution to the shoulders.}
  \label{fig:Particle_morphology_with_den_1}
\end{figure*}

To explore the types of orbits that support shoulders (a more in depth study will be presented in Beraldo e Silva et al., {\it in preparation}), we extract the particles within the shoulder region (\ie{} from inner to outer shoulder edges for both $x<0$ and $x>0$, and $|y|<1$ kpc) for three models at times when the shoulders are well developed. Fig.~\ref{fig:Particle_morphology_with_den_1} shows their surface density in the $(x,y)$-plane at the selection time, and at a later time (but when the shoulders are still present). Although we can see a low density of particles outside of the shoulders (blue areas -- these particles happened to have been located in the shoulder area when the source cut was taken), the distributions overall resemble looped x$_1$ orbits. They have apocentres in the shoulder area, and clearly avoid the centre (we have verified that the shoulder-supporting morphology is not a consequence of our choice of `slit width' in the major-axis cut, $|y|<1$ kpc, by repeating our analysis with significantly larger and smaller slits). This morphology is not seen for particles in cuts outside the shoulders. This is consistent with the evolution of the elongation of orbits of particles destined to be in the shoulders discussed in Section~\ref{ss:shoulder_particle_trapping}.

Tracking the particles further, the lower right pair of plots in Fig.~\ref{fig:Particle_morphology_with_den_1} shows that, with time, the morphology of the shoulder particles `smears out' when viewed in the $(x, y)$-plane. Therefore given enough time, stars initially on shoulder orbits evolve into librating box-like orbits, reducing their support of the shoulder morphology. This in turn requires that, for shoulders to persist long term, the bar must continuously capture additional material onto such looped orbits, and the bar must continue to grow.

\begin{figure*}
  \centering
  \includegraphics[width=\hsize]{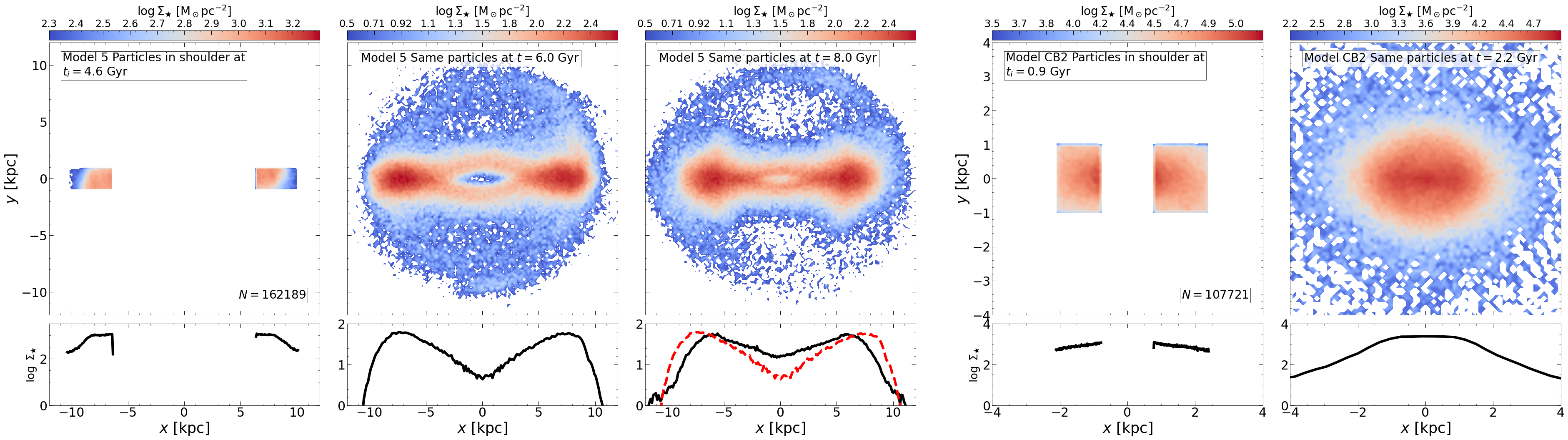}
  \caption{As for Fig.~\ref{fig:Particle_morphology_with_den_1}, but now showing (first three columns) the effect of second buckling and (last two columns) the case of a very weak shoulder. First column: particles in the shoulder at $t_i$ for model 5 (after first buckling). Second and third columns: model 5 shoulder particles at 6 and 8 Gyr, respectively, the latter being 1 Gyr after the second buckling. Fourth and fifth columns: shoulder particles for (non-buckling) model CB2, where the shoulders are extremely weak and we consider them to be transients. In the lower panels we show the major axis density profiles, and in the third column we repeat that for $t=6$ Gyr in red for comparison.}
  \label{fig:Particle_morphology_with_den_2}
\end{figure*}

In Fig.~\ref{fig:Particle_morphology_with_den_2}, we show particles in the shoulders from model 5 at a time before the second buckling (second column) and the same particles after the shoulders have been destroyed by the second buckling (third column). In the third column we also plot (dashed red line) the major axis density from the second column for comparison. We see hints of changes in morphology; the particles do not avoid the centre as much, have a more diffuse, `box-like' shape, and are less extended along $y$ for $|x|\leq3 \kpc$. They are also less radially extended. The mass `smears out' both towards the centre and, at the end of the loops, along the $y$ direction. So the density profile becomes more uniform along the major axis, for particles which were previously concentrated in the loops (dashed red line). The second buckling drives the x$_1$ orbits out of the plane, and this also changes their projected morphology \citep[e.g. Figure~8 in ][]{lokas_2019}, diluting the overdensity. Since the second buckling prevents the bar from growing (temporarily in some models), and therefore renders it unable to capture additional particles from the periphery of the disc into the looped orbits, this transformation weakens and ultimately dissolves the shoulders, returning the bar profile to exponential. Furthermore, as discussed above, $R_\mathrm{CR}$ increases in time, so a relative lack of material outside $R_\mathrm{CR}$ after the second buckling may also contribute to the inability to quickly reform shoulders.

The last two columns of Fig.~\ref{fig:Particle_morphology_with_den_2} show particles in model CB2's very weak shoulders at 0.9 Gyr (we consider these to be transients). We see a diffuse shape and a reduced extension along $y$ for $|x|\leq3$ kpc at 2.2 Gyr. The plots beneath each $(x,y)$-plane panel show the major axis profiles for the selected particles and confirm this visual impression. Weak or dissolved shoulders appear to be supported by more box-like, diffuse orbits.

In their study of orbital support of bars, \citet{smirnov+2021} found that as the central concentration of their models increased, so did the percentage of x$_1$ orbits. We note that the B models reach a higher maximum shoulder strength on average than the WNB models (median $\mathcal{S}_\mathrm{max}$ = 0.25 and 0.16, respectively). Recall that the WNB models are more centrally concentrated than the B models. This would intuitively lead one to expect stronger, not weaker shoulders in the WNB models, in contradiction to our findings. However, the authors do not distinguish between x$_1$ orbits in general, and those having loops, so a direct comparison is difficult.

\section{Discussion}
\label{s:discussion}

\subsection{Shoulders as a manifestation of bar growth}
\label{ss:shoulders_manif_bar_growth}

A key insight of this work is that shoulders are a manifestation of the bar's secular growth. The simulations with prograde spinning haloes (Fig.~\ref{fig:NB_Models_Evolution}) show that shoulders do not form if the bar is unable to grow. Since bars do not form with shoulders in place, shoulders are not part of the bar instability itself, but rather appear if the bar grows. Section~\ref{ss:outer_edge_location} shows that shoulders end just outside the bar radius, with a relatively small variation in $R_\mathrm{sh}/R_\mathrm{bar}$ ($\sigma \sim 5\%$ of the mean), showing that the shoulder edge tracks the end of the bar rather closely.

Fig.~\ref{fig:Shoulder_param_evol} shows that shoulders typically become flatter as they evolve. This is not a general rule, however: models 3 and 4 undergo periods of weakening, and model PBG2 shows no strong growth trend. Furthermore, the data include episodes of steepening (\ie{} becoming less shoulder-like) during secondary buckling and periods of spiral interference. Fig.~\ref{fig:Bar_correls} shows that for many models, in general the stronger and longer the bar, the stronger and flatter becomes the shoulder. This is consistent with \citet{kim+2015} who found that longer bars tend to show flatter profiles (defined by the S\'ersic index of the bar profile) in their sample of 144 face-on barred galaxies. 

However, Figs.~\ref{fig:Shoulder_param_evol} and \ref{fig:Bar_correls} also show that it is not possible to use the observed profile flatness alone to determine the age of a bar, despite the general evolutionary trend towards flattening. The rate at which the slope flattens varies considerably between the models, and the decrease in slope is not monotonic in time (Fig.~\ref{fig:Shoulder_param_evol}) or as the bar strengthens (Fig.~\ref{fig:Bar_correls}). Furthermore, the dissolution of shoulders via secondary buckling and the profile's subsequent return to exponential (grey markers in Figs.~\ref{fig:Shoulder_param_evol} and \ref{fig:Bar_correls}) bolsters the argument that exponential profiles are not necessarily indicators of young bars. We conclude that the flatness of the bar's profile cannot be used as a chronometer, as suggested by \citet{kim+2015}.

Buckling is preceded by a reduction in $\beta=\sigma_z/\sigma_R$, the ratio of the stellar vertical to radial velocity dispersions, to $\sim0.5$ \citep{sellwood96}. We have confirmed that $\beta$ along the bar's major axis declines before the first buckling. In our models, as was seen in simulations by \citet{lokas_2019}, the trigger value is closer to $0.6$. Buckling is followed immediately by a rise in $\beta$ in the buckled region. As the bar continues to grow post buckling, an increase in anisotropy (a renewed reduction in $\beta$) is seen in the shoulder region, as orbits of newly trapped particles become radially elongated (Figs.~\ref{fig:absy_over_absx}, \ref{fig:Particle_morphology_with_den_1}). The increase is focused in the shoulder region since this is the location of highest $\sigma_R$. For the models which undergo a second buckling, $\beta$ eventually reaches $\sim0.6$ in the shoulders once more, leading to buckling in this region, which destroys the shoulders (followed again by an immediate rise in $\beta$). So bar growth triggers both shoulder formation and eventually secondary buckling.  

Persistent shoulders are often, but not always, accompanied by a BP. So a BP can be present without accompanying shoulders, perhaps for them only to emerge after a considerable time (\eg{} model T1 which forms a BP $\sim 2$ Gyr before shoulders appear). In other models, particularly many of the WNB ones, shoulders appear before a BP. The spread in the ratio of shoulder edge to BP radius is twice that with the respect to the bar radius (Section~\ref{ss:outer_edge_location}); thus shoulders track the bar radius rather than the BP bulge radius. It appears that both shoulders and BPs -- although both trace bar growth -- form independently.

\subsection {Orbital support}
\label{ss:orbital_support}

x$_1$ orbits are those primarily responsible for supporting the bar and have been the subject of many studies \citep[see][]{cont_pap_80, sellwood_wilkinson93, binney_tremaine08}. Fig.~\ref{fig:Particle_morphology_with_den_1} suggests that stars lingering in the looped region of x$_1$ orbits are responsible for the shoulders. \citet{contopoulos_1988} found that x$_1$ orbits with loops appeared when the potential was strongly barred, and \citet{athanassoula92} found the same in her models with high quadrupole moment. \citet{valluri+16} discuss examples of x$_1$ orbits, one or two of which qualitatively resemble the shoulder-supporting morphology (their Fig.~4). They are not part of the backbone of the BP structure, being located near the ends of the bar \citep{gajda+2016, parul+2020}, which is near the outer edge of the shoulder structure (Fig.~\ref{fig:quantification}). Hence we expect BPs and shoulders to develop independently despite them both being impacted by bar growth.

As we have shown in Fig.~\ref{fig:Particle_morphology_with_den_2}, shoulder dissolution appears to transform these orbits, possibly into box orbits librating about the x$_1$ orbits. The buckling prevents the bar from growing (temporarily), and therefore renders it unable to capture additional particles into resonant looped x$_1$ orbits. Once the orbits already in the shoulder have evolved into more box-like orbits, the shoulder morphology is no longer supported and the bar profile becomes exponential once more, until fresh stars are trapped onto x$_1$ orbits.

We also note that the location of the inner limit of the shoulder in almost all models moves outwards as the bar evolves (Figs.~\ref{fig:B_Models_Evolution}, \ref{fig:NB_Models_Evolution}). This also supports the notion that the particles trapped into shoulder orbits over time evolve away from looped x$_1$ orbits, into more uniform box-like orbits \citep{lokas_2019}, thus eliminating the overdensity in that part of the bar as it continues to grow outwards.

\subsection{Summary}
\label{ss: summary}

We have used isolated galaxy simulations (16 collisionless $N$-body and three with gas and star formation) to study the outer regions of galactic bars, where the surface density profile along the bar major axis becomes shallow (or `flat') and then breaks to a steep falloff, a pattern we term `shoulders'. Our main results are:

\begin{enumerate}

\item Shoulders form as part of the bar's secular evolution -- they are a sign of a growing bar. They are not present when a bar first forms, and do not subsequently appear if a bar does not grow after formation (see Section~\ref{ss:shoulders_manif_bar_growth}).

\item In many models, the strength and flatness of the shoulders increase as the bar evolves (although not monotonically). Most of our models are consistent with the observational findings of \citet{kim+2015} that stronger and longer bars have flatter profiles (see Section~\ref{ss:sh_quant_properties}).

\item Shoulders often -- but not always -- appear alongside a BP. Some models take a considerable time after BP formation before shoulders emerge and in some models, shoulders appear before a BP, so they are independent tracers of a growing bar. In non-buckling models, stronger and radially larger BP bulges are accompanied by stronger, flatter shoulders (see Section~\ref{sh_correlations}).

\item Secondary buckling dissolves shoulders, either temporarily or longer term (depending on how effectively the bar captures additional material afterwards), returning the bar to an exponential profile without significantly weakening it. Strong spirals perturbing the bar can have the same effect. A thick disc can mask shoulders owing to significant mass being present at large heights. This destruction, the large variety of flattening rates of the bar as it evolves, and the fact that the shoulder growth is not monotonic, means that it is not possible to use the flatness of a bar's profile in a simple way to determine its age. Notably, an exponential bar profile is not necessarily an indication of a young bar (see Sections \ref{s:dissolution}, \ref{ss:shoulders_manif_bar_growth}).

\item For models with both BPs and shoulders, face-on shoulders are evident in projection, even though a `box$+$spurs' morphology in the ($x,y$)-plane isophotes (a BP bulge indicator) may not be. The shoulder slope is strongly affected by projection, particularly at $i\gtrsim 45\degrees$. Caution is therefore urged when observing flat bar profiles in projection (see Section~\ref{s:implications_for_observations}).

\item We showed evidence hinting that shoulders are the manifestation of particles being trapped by the growing bar around looped x$_1$ orbits, where the time spent at apogalacticon results in the overdensity in the shoulder region. In time, these orbits transform, probably into librating boxes, so more material must be trapped for the shoulders to persist.

\item We have verified that our conclusions are consistent with results from the fully self-consistent star forming model, and so are not peculiar to the collisionless models (see Section~\ref{ss:star_forming_model}). 

\end{enumerate}

\section{Acknowledgements}
VPD and LBS are supported by Science and Technology Facilities Council Consolidated grant ST/R000786/1.  DJL was supported for part of this project by UCLan UURIP summer internships in 2020 and 2021. The simulations were run at the High Performance Computing Facility of the University of Central Lancashire
and at the DiRAC Shared Memory Processing system at the University of Cambridge, operated by the COSMOS Project at the Department of Applied Mathematics and Theoretical Physics on behalf of the STFC DiRAC HPC Facility (www.dirac.ac.uk). This equipment was funded by BIS National E-infrastructure capital grant ST/J005673/1, STFC capital grant ST/H008586/1 and STFC DiRAC Operations grant ST/K00333X/1. DiRAC is part of the National E-Infrastructure. We thank the anonymous referee for a careful reading of the manuscript, and for suggestions which improved the paper.

The analysis in this paper made use of the \texttt{PYTHON} packages  \texttt{NUMPY, PYNBODY} and \texttt{SCIPY} \citep{numpy, pynbody,scipy}. Figures were produced using the \texttt{PYTHON} package \texttt{MATPLOTLIB} \citep{matplotlib}.

\section*{Data availability}

The simulations used in this paper may be shared on reasonable request to V.P.D. (vpdebattista@gmail.com).



\bibliographystyle{mnras}
\bibliography{shoulders_main.bbl} 








\bsp	
\label{lastpage}
\end{document}